\documentclass[aps,pre,showkeys,showpacs,a4paper,floatfix,amsmath,amssymb,superscriptaddress]{revtex4}
\usepackage{natbib}
\usepackage{dcolumn}
\usepackage{graphicx}
\usepackage{epsfig}

\begin{document}
\title{Deterministic event-based simulation of quantum interference
}

\author{K. De Raedt}
\email{deraedt@cs.rug.nl}
\affiliation{Department of Computer Science,
University of Groningen, Blauwborgje 3,
NL-9747 AC Groningen, The Netherlands}
\author{H. De Raedt}
\email{deraedt@phys.rug.nl}
\homepage{http://www.compphys.org}
\affiliation{Department of Applied Physics,
Materials Science Centre, University of Groningen, Nijenborgh 4,
NL-9747 AG Groningen, The Netherlands
}
\author{K. Michielsen}
\email{kristel@phys.rug.nl}
\affiliation{Department of Applied Physics,
Materials Science Centre, University of Groningen, Nijenborgh 4,
NL-9747 AG Groningen, The Netherlands
}
\begin{abstract}
We propose and analyse simple deterministic algorithms that
can be used to construct machines that have primitive learning capabilities.
We demonstrate that locally connected networks of these machines
can be used to perform blind classification on an event-by-event basis,
without storing the information of the individual events.
We also demonstrate that properly designed networks of these machines exhibit
behavior that is usually only attributed to quantum systems.
We present networks that simulate quantum interference on an event-by-event basis.
In particular we show that by using simple geometry and the
learning capabilities of the machines it becomes possible to simulate
single-photon interference in a Mach-Zehnder interferometer.
The interference pattern generated by the network of deterministic
learning machines is in perfect agreement with the quantum theoretical result
for the single-photon Mach-Zehnder interferometer.
To illustrate that networks of these machines
are indeed capable of simulating quantum interference
we simulate, event-by-event, a setup involving two chained Mach-Zehnder interferometers.
We show that also in this case the simulation results agree with quantum theory.
\end{abstract}
\date{\today}

\maketitle

\def\ORDER#1{\hbox{${\cal O}(#1)$}}
\def\BRA#1{\langle #1 \vert}
\def\KET#1{\vert #1 \rangle}
\def\EXPECT#1{\langle #1 \rangle}
\def\BRACKET#1#2{\langle #1 \vert #2 \rangle}
\def\hbar{{\mathchar'26\mskip-9muh}}
\def\mod{{\mathop{\hbox{mod}}}}
\def\CNOT{{\mathop{\hbox{CNOT}}}}
\def\Tr{{\mathop{\hbox{Tr}}}}
\def\bPsi{{\mathbf{\Psi}}}
\def\bPhi{{\mathbf{\Phi}}}
\def\bzero{{\mathbf{0}}}
\def\Eq#1{(\ref{#1})}

\def\DLM{DLM}
\def\DLMS{DLMs}

\section{Introduction}\label{intro}

Computer simulation is widely regarded as complementary to theory and experiment~\cite{LAND00}.
At present there are only a few physical phenomena that cannot be simulated on a computer. 
One such exception is the double-slit experiment with single electrons, as carried out by
Tonomura and his co-workers~\cite{TONO98}.
This experiment is carried out in such a way that at any given time, only one electron travels
from the source to the detector~\cite{PerfectExperiments}.
Only after a substantial (approximately 50000) amount of
electrons have been detected an interference pattern emerges~\cite{TONO98}.
This interference pattern is described by quantum theory.
We use the term ``quantum theory'' for the mathematical formalism
that gives us a set of algorithms to compute the probability
for observing a particular event~\cite{HOME97,KAMP88,QuantumTheory}.
Of course, the quantum-mechanics textbook example~\cite{FEYN65,BALL03} of a double-slit
can be simulated on a computer by solving the time-dependent Schr{\"o}dinger equation
for a wave packet impinging on the double slit~\cite{RAED96,QMvideo}.
Alternatively, in order to obtain the observed interference pattern
we could simply use random numbers to generate events according to the probability distribution
that is obtained by solving the time-independent Schr{\"o}dinger equation.
However, that is not what we mean when we say that
the physical phenomenon cannot be simulated on a computer.
The point is that it is not known how to simulate,
event-by-event, the experimental observation that the
interference pattern appears only after a considerable
number of events have been recorded on the detector.
Quantum theory does not describe the individual events, e.g.
the arrival of a single electron at a particular position 
on the detection screen~\cite{TONO98,HOME97,FEYN65,BALL03}.
Reconciling the mathematical formalism (that does not describe single events) with
the experimental fact that each observation yields a definite
outcome is often referred to as the quantum measurement paradox
and is the central, most fundamental problem in the foundation
of quantum theory~\cite{FEYN65,HOME97,PENR90}.

If computer simulation is indeed a third methodology 
to model physical phenomena it should be possible to simulate
experiments such as the two-slit experiment on an event-by-event basis.
In view of the fundamental problem alluded to above there
is little hope that we can find a simulation algorithm 
within the framework of quantum theory.
However, if we think of quantum theory as a set of algorithms to compute probability
distributions there is nothing that prevents us from stepping outside the framework
that quantum theory provides.
Therefore we may formulate the physical processes in terms of events, messages,
and algorithms that process these events and messages,
and try to invent algorithms that simulate the physical processes.
Obviously, to make progress along this line of thought, it makes sense not
to tackle the double-slit experiment directly but to simplify the problem
while retaining the fundamental problem that we aim to solve.

The main objective of the research reported in this paper is to answer the question:
``Can we simulate the single-photon beam splitter and Mach-Zehnder interferometer
experiments of Grangier et al.~\cite{GRAN86} on an event-by-event basis?''.
These experiments display the same fundamental problem as the single-electron
double-slit experiments but are significantly easier to describe in terms of algorithms.
The main results of our research are that we can give an affirmative answer
to the above question by using algorithms that have a primitive form of
learning capability and that the simulation approach that we propose 
can be used to simulate other quantum systems (including the double-slit experiment) as well. 

In Section~\ref{ILLU} we introduce the basic concepts for constructing event-based,
deterministic learning machines (\DLMS).
An essential property of these machines is that they process input event after
input event and do not store information about individual events.
A \DLM\ can discover relations between input events (if there are any)
and responds by sending its acquired knowledge 
in the form of another event (carrying a message) through one of its output channels.
By connecting an output channel to the input channel of another \DLM\ we can build
networks of \DLMS.
As the input of a network receives an event,
the corresponding message is routed through the network while it is being
processed and eventually a message appears at one of the outputs.
At any given time during the processing,
there is only one input-output connection in the network that is actually carrying a message.
The \DLMS\ process the messages in a sequential manner and communicate with each other
by message passing.
There is no other form of communication between different \DLMS.
Although networks of \DLMS\ can be viewed as networks that are capable of unsupervised learning,
there have very little in common with neural networks~\cite{HAYK99}.
The first \DLM\ described in Section~\ref{ILLU} is equivalent to a
standard linear adaptive filter~\cite{HAYK86}
but the \DLMS\ that we actually use for our applications
do not fall into this class of algorithms.

In Section~\ref{NDIM} we generalize the ideas of Section~\ref{ILLU} and
construct a \DLM\ which groups $K$-dimensional data in two classes
on an event-by-event basis, i.e., without using memory to store the whole data set. 
We demonstrate that this \DLM\ is capable of detecting time-dependent trends in the data
and performs blind classification.
This example shows that \DLMS\ can be used to solve problems
that have no relation to quantum physics. 

In Section~\ref{QI} we show how to construct \DLM-networks that generate output patterns
that are usually thought of as being of quantum mechanical origin.
We first build a \DLM-network that simulates photons passing through a polarizer
and show that quantum theory describes the output of this deterministic,
event-based network.
Then we describe a \DLM-network that simulates a beam splitter
and use this network to build a Mach-Zehnder interferometer
and two chained Mach-Zehnder interferometers.
We demonstrate that quantum theory also describes the
behavior of these networks.

Quantum theory gives us a recipe to compute the frequency of events but does
not predict the order in which the events will be observed~\cite{BALL03,HOME97}.
In genuine experiments the detection of events appears to be random~\cite{GRAN86,TONO98},
in a sense which, as far as we know, has not been studied systematically.
In our simulation approach, this apparent randomness can be accounted
for by a marginal modification of the \DLMS, as explained in Section~\ref{SLM}.
This modification does not change the deterministic character of the learning process.
It merely randomizes the order in which the \DLMS\ activate their output channels.

A summary and outlook is given in Section~\ref{SUMM}.

\def\DLM{machine}
\def\DLMS{machines}
\section{Deterministic learning machines}\label{ILLU}
\subsection{Learning points on the real axis}\label{ILLUa}

We consider a \DLM\ that has one input and two output channels labeled by $\pm1$ (see Fig.~1).
The internal state of the \DLM\ after processing the $n$-th input event ($n=0,1,\ldots$)
is uniquely defined by the real variable $x_n$.
At the next event $n+1$ the \DLM\ receives as input a real number $y_{n+1}$.
For simplicity, but without loss of generality,
we assume that $y_{n+1}\in[-1,1]$.
The \DLM\ responds by sending a message containing $y_{n+1}$
through one of the two output channels $\Delta_{n+1}=\pm1$.
The \DLM\ selects the output channel $\Delta_{n+1}=+1$ or $\Delta_{n+1}=-1$ by
minimizing the cost function $C(\Delta_{n+1})$ defined by

\begin{equation}
C(\Delta_{n+1})=|y_{n+1}-x_n - (1-\alpha)\Delta_{n+1}|y_{n+1}-x_n||
,
\label{ILLU1}
\end{equation}
updates its internal state according to the rule

\begin{equation}
x_{n+1}=x_n + (1-\alpha)\Delta_{n+1}|y_{n+1}-x_n|
,
\label{ILLU2}
\end{equation}
and sends a message with the input value $y_{n+1}$
on the selected output channel $\Delta_{n+1}$.
The parameter $0<\alpha<1$ that enters Eqs.~\Eq{ILLU1} and \Eq{ILLU2}
controls the decision process.
For simplicity we assume that $\alpha$ is fixed
during the operation of the machine.

It is easy to see that
$\Delta_{n+1}=+1$ if $x_n\le y_{n+1}$
and $\Delta_{n+1}=-1$ if $x_n>y_{n+1}$.
Thus, for this particular \DLM\ we have

\begin{equation}
\Delta_{n+1}=\frac{y_{n+1}-x_n}{|y_{n+1}-x_n|}
.
\label{ILLU3}
\end{equation}
Hence the update rule Eq.~\Eq{ILLU2} can be written as
the familiar recursion

\begin{equation}
x_{n+1}=\alpha x_n + (1-\alpha)y_{n+1}
.
\label{ILLU4}
\end{equation}
The solution of Eq.~\Eq{ILLU4} reads

\begin{equation}
x_{n}=\alpha^n x_0 + (1-\alpha)\sum_{i=0}^{n-1}\alpha^{n-1-i}y_{i+1}
,
\label{ILLU5}
\end{equation}
where $x_0$ denotes the initial value of the internal variable.

As an illustration of how this \DLM\ learns,
we consider the most simple example where $y_{n+1}=y$ for all $n\ge0$.
Then from Eq.~\Eq{ILLU5} we find that

\begin{equation}
x_{n}=\alpha^n x_0 + (1-\alpha^n)y
.
\label{ILLU6}
\end{equation}
As $0<\alpha<1$, we conclude that $\lim_{n\rightarrow\infty} x_n=y$.
Thus the \DLM\ ``learns'' the value of the input variable $y$.
From Eq.~\Eq{ILLU4} it follows that $x_n\le y$ ($x_n\ge y$)
implies $x_{n+1}\le y$ ($x_{n+1}\ge y$).
Hence $x_n$ approaches $y$ monotonically
(and $\Delta_n$ is the same for all $n$).
Therefore, if $y_n=y$, the \DLM\ always sends
the value of $y_n$ through the same output channel.

\begin{figure}[t]
\begin{center}
\setlength{\unitlength}{1cm}
\begin{picture}(8,6)
\put(-5,1){\includegraphics[width=9cm]{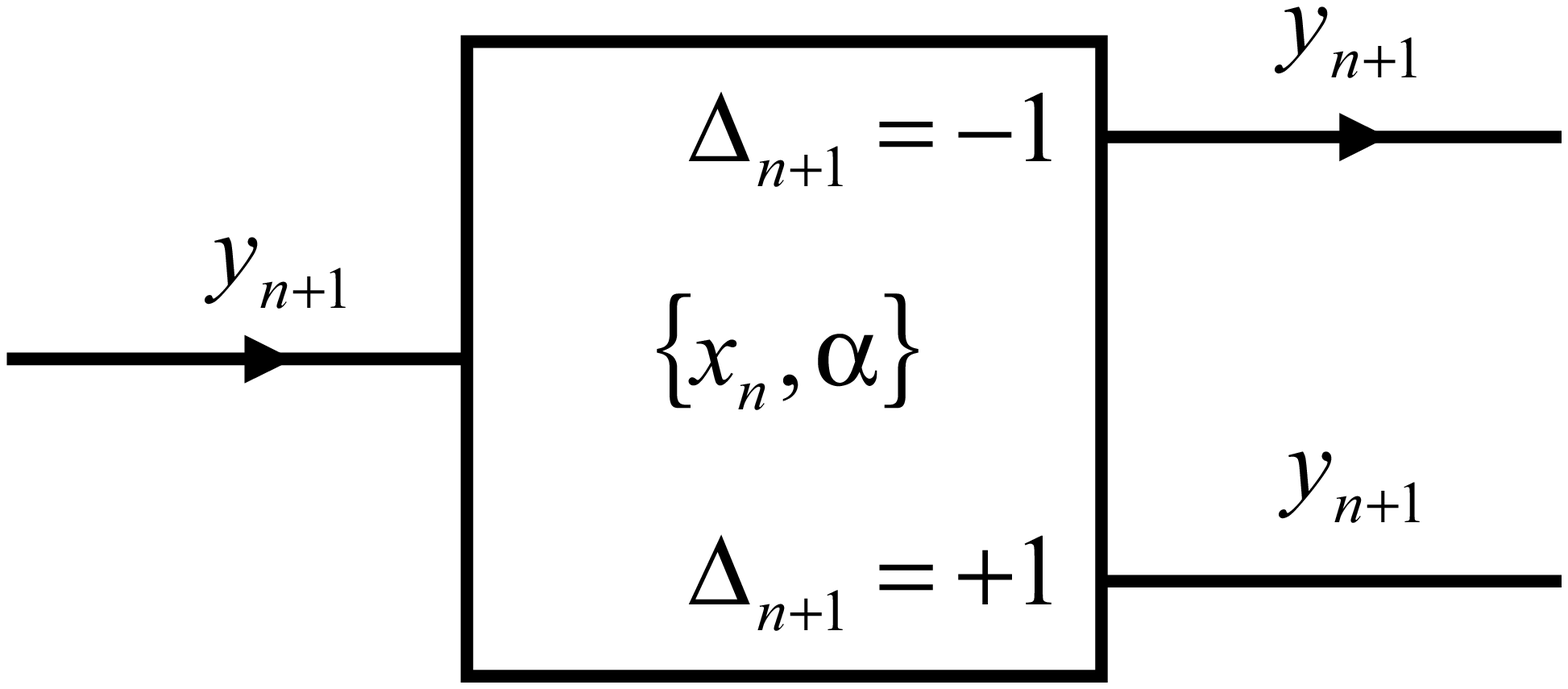}}
\put(4,0){\includegraphics[width=9cm]{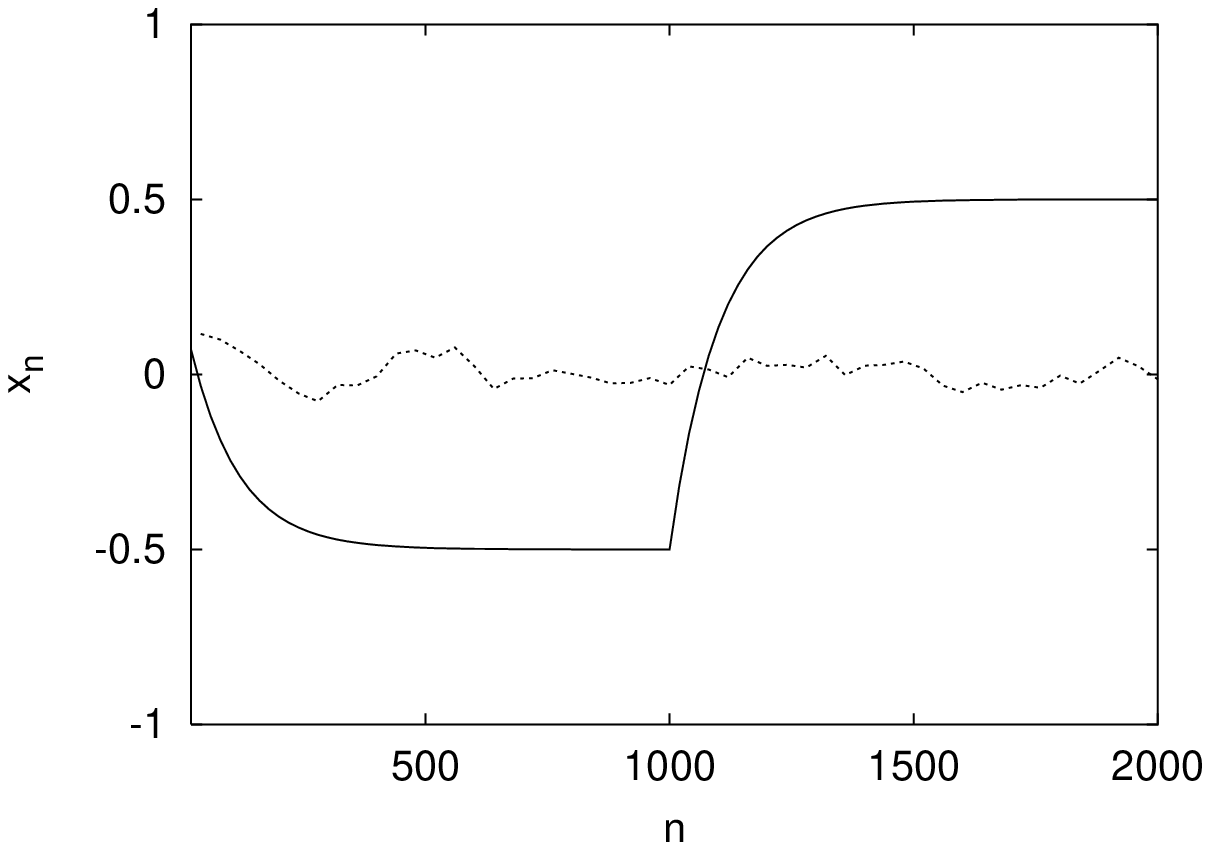}}
\end{picture}
\caption{%
Left:
Schematic representation of the \DLM\ that
responds to the input $y_{n+1}$ by passing the input
to one of the two output channels $\Delta_{n+1}=\pm1$.
The value of $\Delta_{n+1}$ depends on the current
state of the \DLM, encoded in the variable $x_{n}$,
the input $y_{n+1}$, and the update rule Eq.~\Eq{ILLU2}
in which $\alpha$ appears as a control parameter.
Right:
Evolution of the internal variable $x_{n}$
as a function of the number of events $n$.
Solid line: $y_{n+1}=-0.5$ for $n=1,\ldots,1000$
and $y_{n+1}=0.5$ for $n=1001,\ldots,2000$;
Dashed line: Random sequence of $y_{n+1}=\pm 0.5$.
}
\label{exam01}
\end{center}
\end{figure}

A distinct feature of this machine is its ability to adapt
to changes in the input pattern. We illustrate this
important property by two examples.
Let $y_n = -0.5$ for $1\le n\le1000$
and $y_n = 0.5$ for $1000<n\le2000$.
During the first 1000 events the machine will learn $-0.5$.
After 1000 events only $0.5$ is being presented as input.
Then, the machine ``forgets'' $-0.5$ and learns $0.5$
as shown in the right panel of Fig.~\ref{exam01}.
In this simulation $\alpha=0.99$.
Alternatively, if $y_n$ is a random sequence of $\pm 0.5$ (each with
the same probability) the machine has to learn $-0.5$ and $0.5$ simultaneously.
Because of this it cannot ``forget'' and it ends up oscillating
around the mean of the input values (zero in this example)
as illustrated in the right panel of Fig.~\ref{exam01}.
Let us now assume that our machine has reached this oscillating state.
All input events $y_n=0.5$ give $\Delta_n=+1$ and hence the machine sends $0.5$
over the $+1$ channel.
A second machine attached to this channel only receives $0.5$ events and will learn $0.5$.
This suggests that a network of these machines can be used as an adaptive classifier.

\begin{figure}[t]
\begin{center}
\setlength{\unitlength}{1cm}
\begin{picture}(8,12)
\put(-4.5,3){\includegraphics[width=8cm]{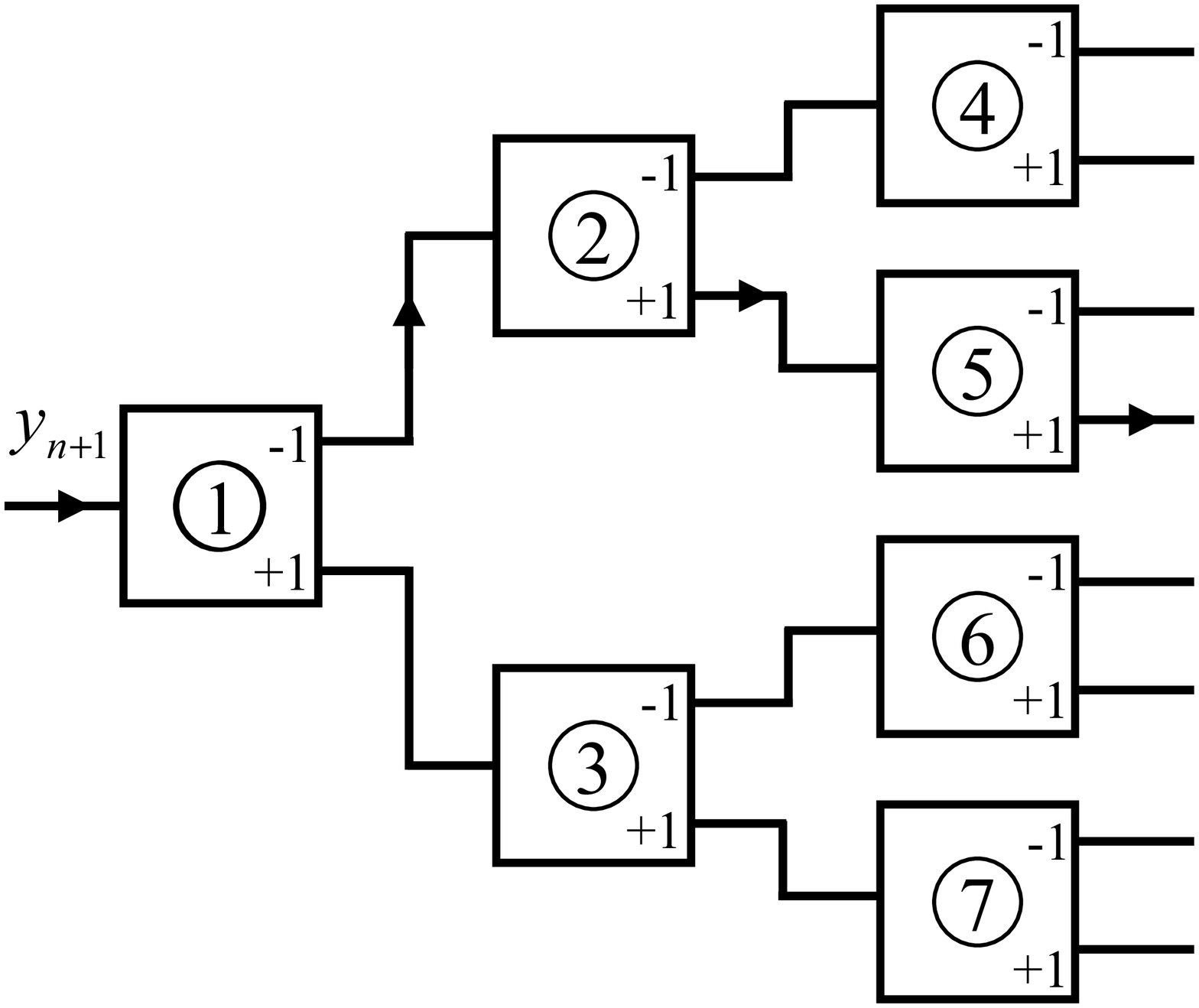}}
\put(4,6){\includegraphics[width=9cm]{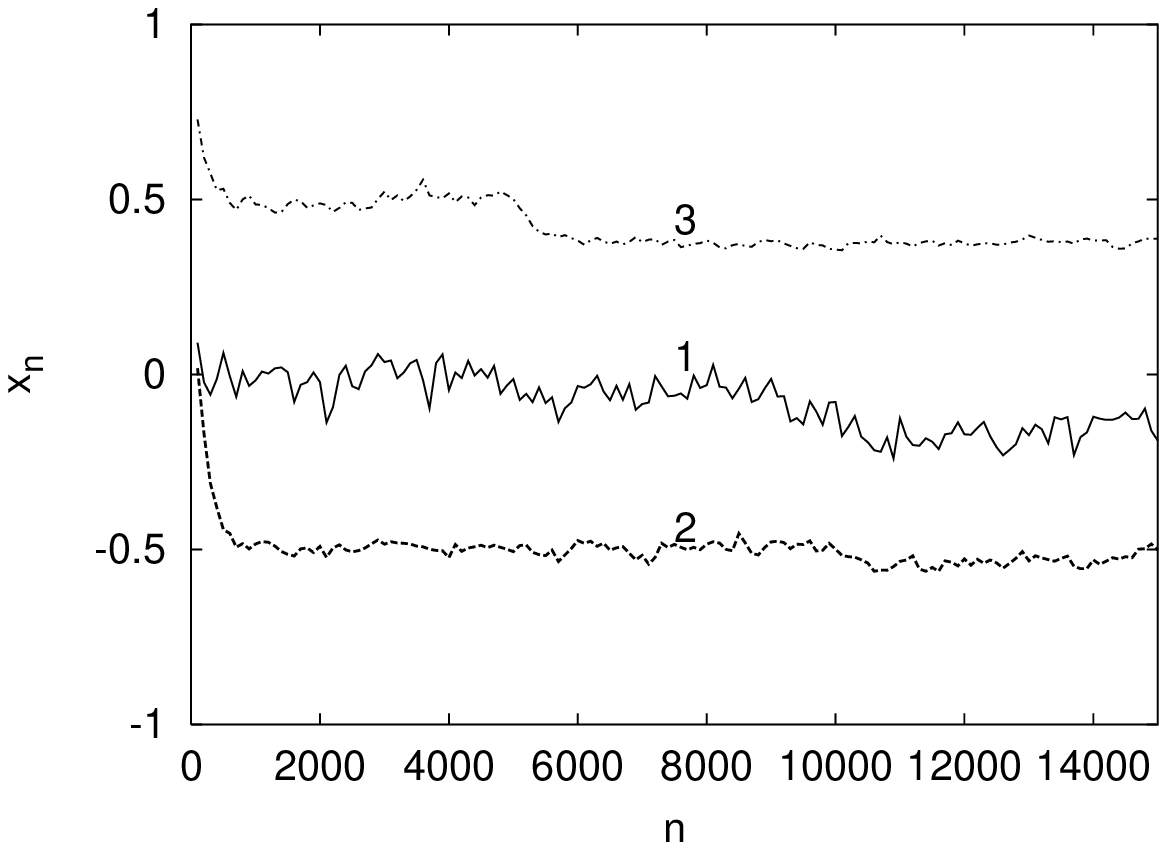}}
\put(4,0){\includegraphics[width=9cm]{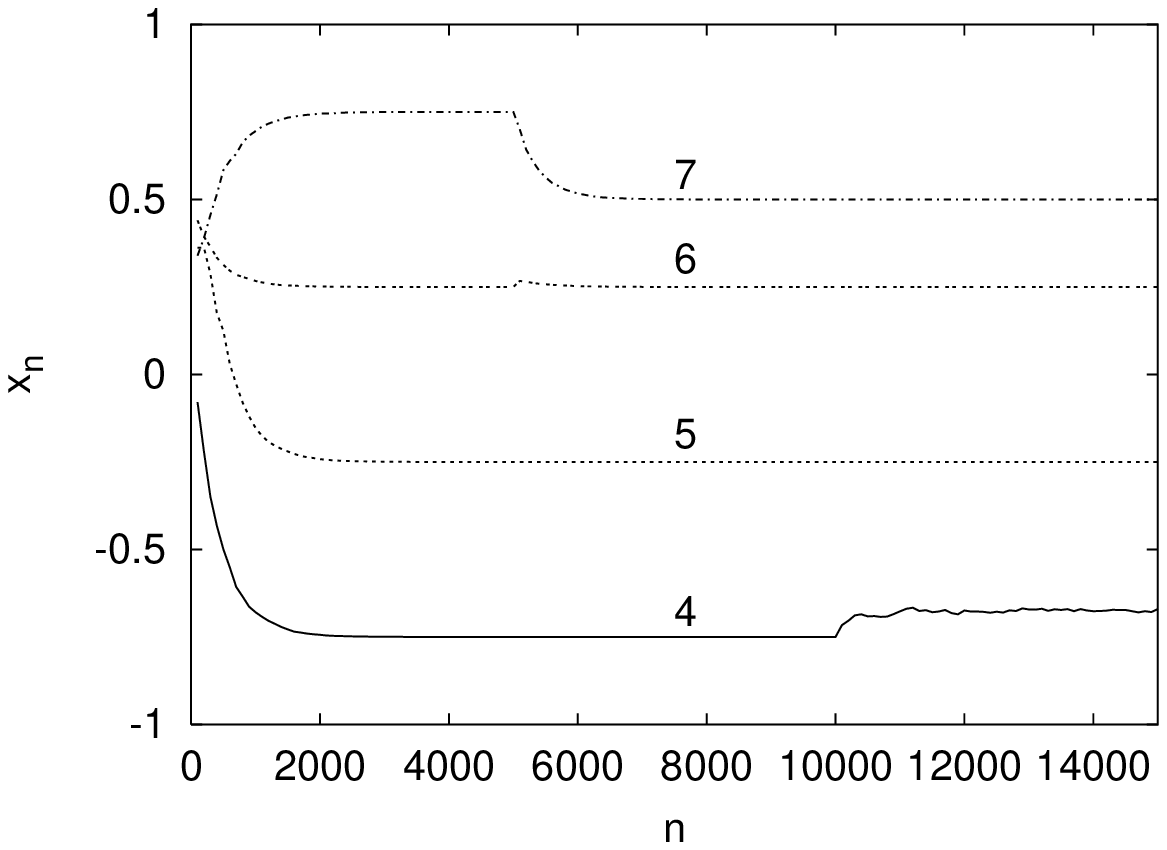}}
\end{picture}
\caption{%
Left:
Diagram of the three-level \DLM\ that adaptively classifies
the input data $y_{n+1}$.
Right:
Evolution of the internal variables $x_{n}$
of the \DLMS\ as a function of the number of events $n$.
The machine number is used to label the corresponding line.
Top right: First three \DLMS;
Bottom right: Third-level \DLMS.
}
\label{exam03}
\end{center}
\end{figure}

Consider the network of three layers of \DLMS\ shown in
the left panel of Fig.~\ref{exam03}.
Each machine in the network learns the average of the numbers it receives at its input
channel and sends the numbers which are smaller (larger or equal) than the number it
learned to the -1 (+1) output channel.
In our numerical experiments we set $\alpha=0.99$.
We start with 5000 events of random numbers $y_{n+1} \in \{-0.75,-0.25,0.25,0.75\}$,
each occurring with equal probability.
Machine 1 learns the average (zero in this example)
and sends the negative (positive) $y_{n+1}$ over the $-1$ ($+1$) channel
to the input of machine 2 (3).
Machine 2 (3) learns -0.50 (0.50), as shown in the top right panel of Fig.~\ref{exam03},
and sends -0.75 (0.25) over its -1 output channel and -0.25 (0.75) over its
+1 output channel.
Machines 4 to 7 learn -0.75,-0.25,0.25 and 0.75, respectively,
as shown in the bottom right panel of Fig.~\ref{exam03}.
Each of these machines forwards the received input on its +1 (-1) output channel
if the initial value of its internal variable is smaller (larger)
than the received input value.
Let us now assume that after 5000 events
the input data set changes to $y_{n+1} \in \{-0.75,-0.25,0.25,0.50\}$.
As can be seen from the right panel of Fig.~\ref{exam03},
machines 1, 3 and 7 ``forget'' the number they learned and
replace it by -0.0625, 0.375 and 0.50, respectively.
All other machines are unaffected because they never get 0.50 as input.
After another 5000 events we change the set of input values once more, this time to
$y_{n+1} \in \{-0.60, -0.75,-0.25,0.25,0.50\}$, i.e., we add one element.
Now, machine 1 learns -0.17, machine 2 learns -0.53 and the internal state of
machine 3 remains unchanged.
Machine 4 can now receive two numbers on its input channel, namely -0.75 and -0.60.
As a consequence, machine 4 learns -0.675,
i.e., the average of the two possible input numbers.
Machine 4 puts -0.60 on its +1 output channel and -0.75 on its -1 output channel.
In order for the network to learn all
the numbers of the input set, we would have to attach one extra \DLM\ to
each output channel of machine 4.

\setlength{\unitlength}{1cm}
\begin{figure}[t]
\begin{center}
\begin{picture}(8,6)
\put(-5,0){\includegraphics[width=9cm]{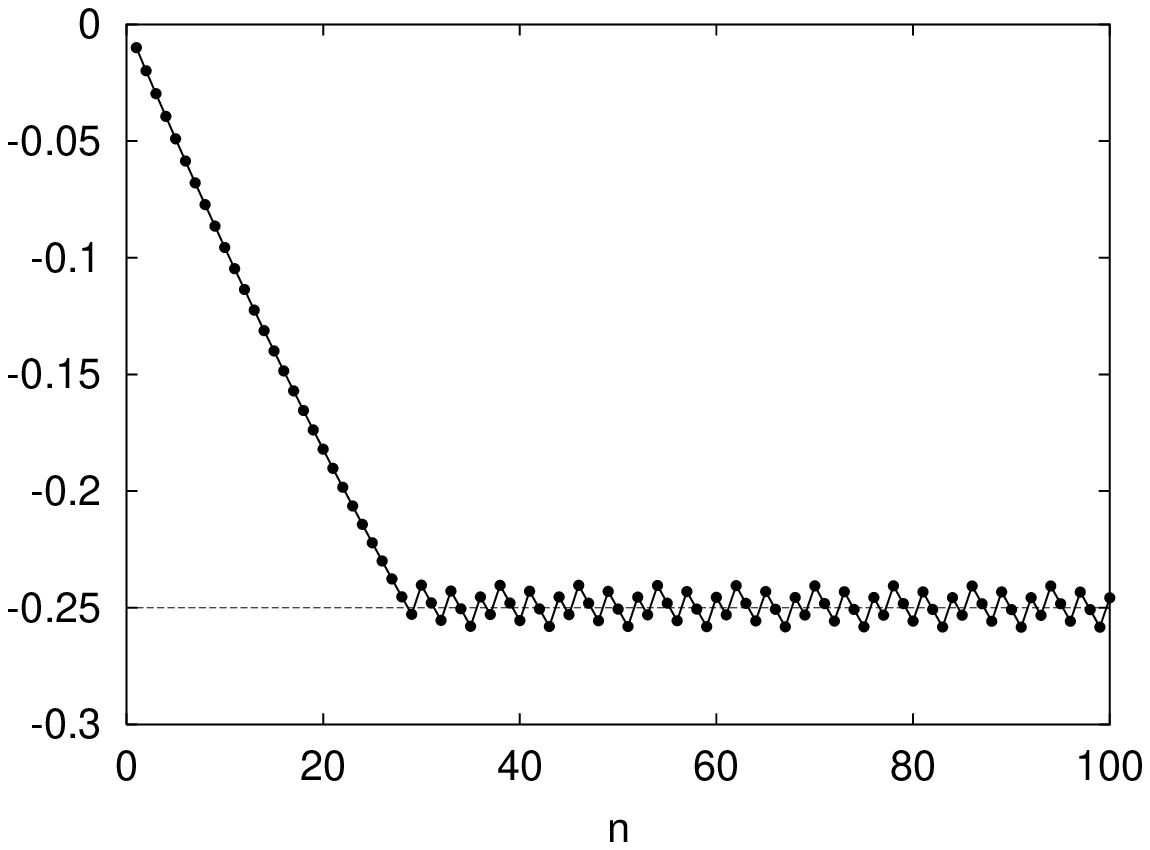}}
\put(4,0){\includegraphics[width=9cm]{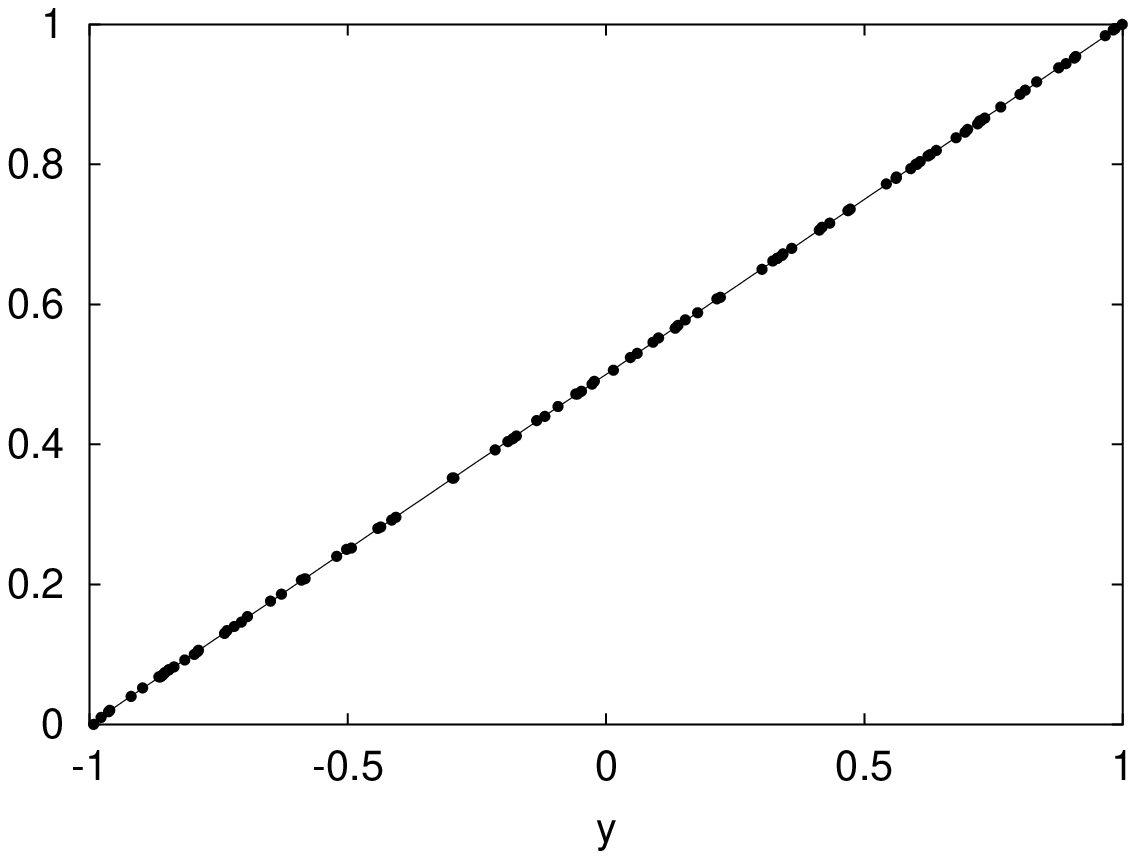}}
\end{picture}
\caption{%
Left:
Time evolution of the internal variable $x_n$ of the
machine defined by Eqs.~\Eq{LINE1} and \Eq{LINE2}.
The input events are $y=-0.25$, $\alpha=0.99$,
and the initial value $x_0=0$.
For $n>30$ the internal variable $x_n$ oscillates about $y$.
For $n>500$ the sequence of increments ($\Delta_{n+1}=+1$)
and decrements ($\Delta_{n+1}=-1$) of $x_n$
repeats itself after 8 events (data not shown).
Lines are guides to the eyes.
Right:
The number of increments of the internal variable
($\Delta_{n+1}=+1$) divided by the total number of events
as a function of the value of the input variable $y$.
Bullets: Each data point is obtained from a
simulation of 1000 events with a fixed, randomly
chosen value of $-1< y <1$, using the last 500 events
to count the number of $\Delta_{n+1}=+1$ events.
Solid line: $(1+y)/2$.
}
\label{figline}
\end{center}
\end{figure}

\subsection{Learning points on a finite interval}\label{ILLUb}

For the \DLM\ defined by Eqs.~\Eq{ILLU1} and \Eq{ILLU2},
formulating the operation of the \DLM\
through the minimization of the difference between
the input and internal variable may seem a little
superfluous and indeed, for this particular machine it is.
However, this formulation is a convenient starting point for
defining machines that can perform more intricate tasks.
For instance, let us make an innocent looking
change to the update rule Eq.~\Eq{ILLU2} by writing

\begin{equation}
x_{n+1}=\alpha x_n + (1-\alpha)\Delta_{n+1}
,
\label{LINE1}
\end{equation}
and replace the cost function Eq.~\Eq{ILLU1}
by the corresponding expression

\begin{equation}
C(\Delta_{n+1})=|y_{n+1}-\alpha x_n - (1-\alpha)\Delta_{n+1}|
.
\label{LINE2}
\end{equation}
For $\Delta_{n+1}=+1$ we have $x_{n+1}=1-\alpha(1-x_n)$
and for $\Delta_{n+1}=-1$ we have $x_{n+1}=-1+\alpha(1+x_n)$.
Therefore, if $0< \alpha < 1$ and $|x_0|\le1$, the internal variable
will always be in the range $[-1,1]$.
At each event the internal variable
either increases by $(1-\alpha)(1-x_n)$ (if $\Delta_{n+1}=+1$)
or decreases by $(1-\alpha)(1+x_n)$ (if $\Delta_{n+1}=-1$).
In both cases this change is always nonzero,
except if $x_n=\pm1$ which can only occur if $y_{n+1}=\pm1$.
The ratio of the step sizes is $(1-x_n)/(1+x_n)$.

The machine defined by Eqs.~\Eq{LINE1} and \Eq{LINE2} behaves differently
from the machine defined by Eqs.~\Eq{ILLU1} and \Eq{ILLU2}.
To see this, it is instructive to consider the case $0\le y_{n+1}=y<1$ for all $n\ge0$
(the case $-1<y_{n+1}=y<0$ can be treated in the same manner).
For concreteness we assume that $-1< x_0<y$.
At the first event, minimization of Eq.~\Eq{LINE2} yields
$\Delta_{1}=+1$ and $x_1=1+\alpha(x_0-1)$.
In other words, the internal variable $x$ moves towards $y$.
As long as $x_n<y$, the \DLM\ selects $\Delta_{n+1}=+1$,
always increasing its internal variable $x_n$.
For some some $n\ge1$ we must have $x_n>y$.
Then, making another move in the positive $x$-direction
allows for two different decisions.
If the error that results is larger than the error that
is obtained by moving in the negative direction
the \DLM\ decides to set $\Delta_{n+1}=-1$.
Otherwise it makes another move in the positive $x$-direction ($\Delta_{n+1}=+1$).
In any case, for some $n>1$ the machine will select $\Delta_{n+1}=-1$.
Note that when this happens, we must have $x_{n+1}<y$ and $\Delta_{n+2}=+1$.
This implies that after this $n$-th event (that we denote by $n_0$)
the internal variable will oscillate (forever) around the input value $y$.
This process is illustrated in Fig.~\ref{figline} (left).

For $m>n_0$ we have $|x_{m+1}-y| \le (1-\alpha)\max(1-y,1+y)$.
Thus, if $0<1-\alpha\ll1$, the amplitude of the oscillations is small.
The \DLM\ ``learns'' the input value $y$ and
the ratio of the increments to decrements is $(1+x_{m+1})/(1-x_{m+1})\approx(1+y)/(1-y)$.
In this stationary regime of oscillating behavior, the number
of times the \DLM\ actives the +1 (-1) channel is given by $(1+y)/2$ ($(1-y)/2$).
The simulation results shown in Fig.~\ref{figline} (right) confirm the correctness of
this analysis.
For a fixed (unknown) value of the input variable,
the rate at which the machine defined by the rules Eqs.~\Eq{LINE1} and \Eq{LINE2}
activates one of its output channels is determined by the value
of its internal variable.
Therefore, this rate reflects the value that the machine has learned
by processing the input events.
Depending on the application, the message that is sent through
the active output channel can contain $x_{n+1}$ or the input value $y_{n+1}$
(there is nothing else that can be send).
Obviously we can make the learning process more precise by increasing $\alpha<1$.
Of course, a larger value of $\alpha$ also results in slower learning: In general it will
take more events for the internal variable to reach the value where it starts to oscillate.

\subsection{Learning points on a circle}\label{CIRC}

In going from the first to the second example of Section~\ref{ILLU}
we changed the update rule such that the variable $x_n$ is constrained
to lie in the interval $[-1,1]$.
We now consider the two-dimensional analogue of the \DLMS\ described in Section~\ref{ILLUb}
for which the internal vector $(x_{1,n},x_{2,n})$
and input vector $(y_{1,n+1},y_{2,n+1})$ represent points on a circle.
This \DLM\ receives as input a sequence of angles $\phi_{n+1}$
defined by

\begin{eqnarray}
\cos \phi_{n+1}&=&\frac{y_{1,n+1}}{\sqrt{y_{1,n+1}^2+y_{2,n+1}^2}},
\nonumber \\
\sin \phi_{n+1}&=&\frac{y_{2,n+1}}{\sqrt{y_{1,n+1}^2+y_{2,n+1}^2}},
\label{NDIMangle}
\end{eqnarray}
and responds by activating one of the two output channels.

For all $n>0$, the update rules are defined by

\begin{eqnarray}
x_{1,n+1}&=&\alpha x_{1,n} + \beta\Theta_{n+1},
\nonumber \\
x_{2,n+1}&=&\alpha x_{2,n} + \beta(1-\Theta_{n+1}),
\label{CIRC1}
\end{eqnarray}
where $\Theta_{n+1}=0,1$ and $0<\alpha<1$.
In order that the internal vector
${\bf x}_{n+1}=(x_{1,n+1},x_{2,n+1})$ stays on the unit circle
we must have

\begin{equation}
\beta= -\alpha[x_{1,n}\Theta_{n+1} +x_{2,n}(1-\Theta_{n+1})]
\pm\sqrt{1-\alpha^2+\alpha^2[x_{1,n}^2\Theta_{n+1} +x_{2,n}^2(1-\Theta_{n+1})]}
.
\label{CIRC2}
\end{equation}
Substitution of Eq.~\Eq{CIRC2} in Eq.~\Eq{CIRC1} gives us
four different rules:

\begin{eqnarray}
x_{2,n+1}&=&s\sqrt{1+\alpha^2(x_{2,n}^2-1)}
\quad,\quad
x_{1,n+1}=\alpha x_{1,n}
\quad\hbox{if } \Theta_{n+1}=0,
\nonumber \\
x_{1,n+1}&=&s\sqrt{1+\alpha^2(x_{1,n}^2-1)}\quad,\quad
x_{2,n+1}=\alpha x_{2,n}\quad\hbox{if } \Theta_{n+1}=1,
\label{CIRC3}
\end{eqnarray}
where $s=\pm 1$ takes care of the fact that
for each choice of $\Theta_{n+1}$, the \DLM\ has to decide between two quadrants.
The cost function is defined by

\begin{equation}
C= -(x_{1,n+1}y_{1,n+1}+x_{2,n+1}y_{2,n+1})
.
\label{CIRC4}
\end{equation}
Obviously, the cost function Eq.~\Eq{CIRC4} is nothing but the
inner product of the vectors ${\bf x}_{n+1}$ and ${\bf y}_{n+1}$.
The new internal state itself is determined by calculating the cost Eq.~\Eq{CIRC4}
for each of the four candidate update rules listed in Eq.~\Eq{CIRC3}
and selecting the rule that yields the minimum cost.
Note that the minimum of the cost function Eq.~\Eq{CIRC4} does not depend on
the length of the vector of input variables $(y_{1,n+1},y_{2,n+1})$.
From Eq.~\Eq{CIRC3} it follows that if $\Theta_{n+1}=0$.
the value of $x_{1,n+1}$ is obtained by rescaling
of $x_{1,n}$ and $x_{2,n+1}$ is adjusted such that
$x_{1,n+1}^2+x_{2,n+1}^2=1$. For $\Theta_{n+1}=1$
we interchange the role of the first and second element of ${\bf x}_{n+1}$.

\setlength{\unitlength}{1cm}
\begin{figure*}[t]
\begin{center}
\begin{picture}(8,6)
\put(-5,0){\includegraphics[width=9cm]{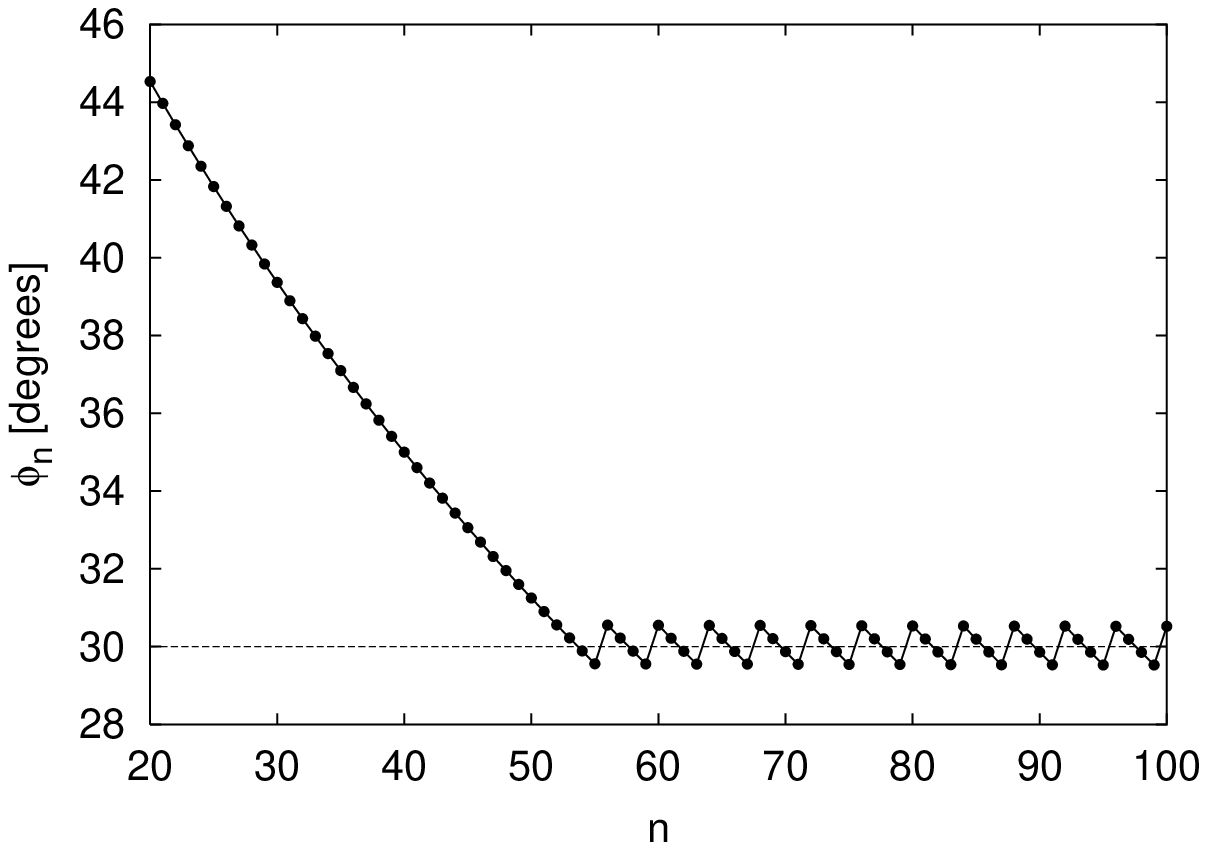}}
\put(4,0){\includegraphics[width=9cm]{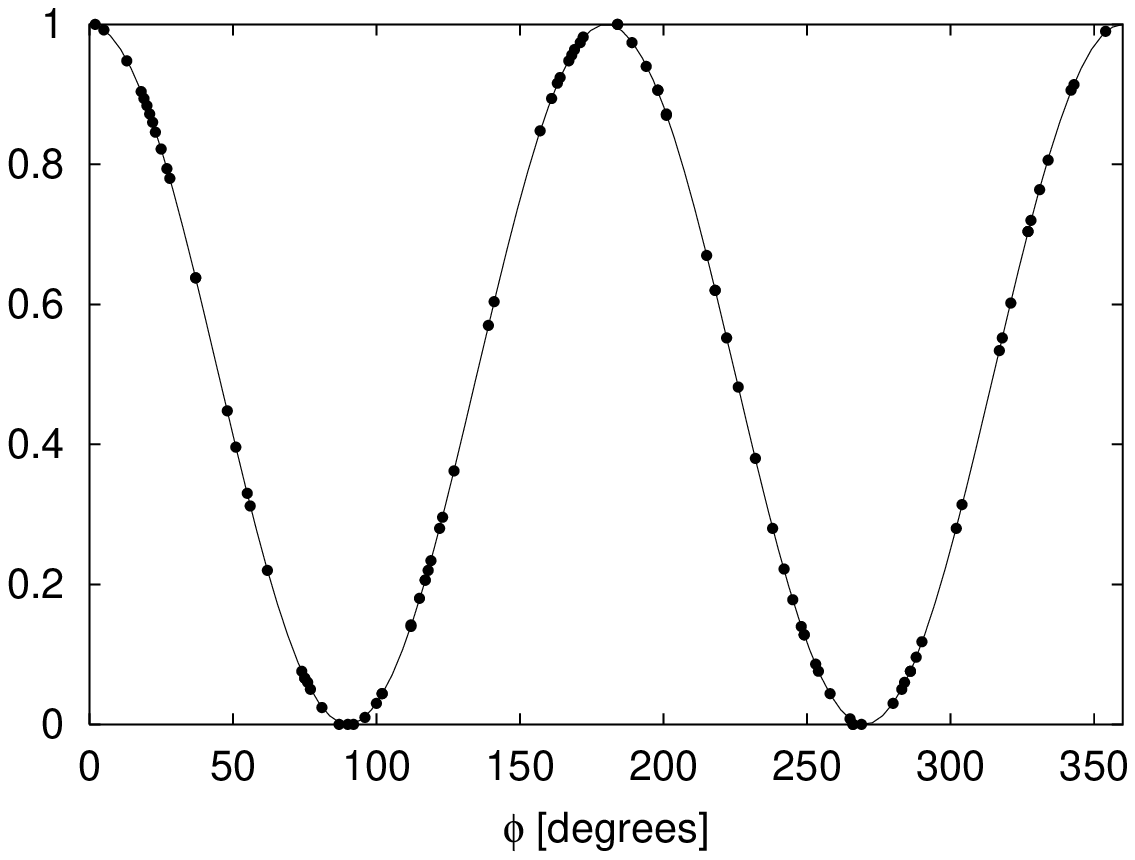}}
\end{picture}
\caption{%
Left:
Time evolution of the angle representing the internal vector ${\bf x}_{n}$
of the \DLM\ defined by Eqs.~\Eq{CIRC3} and \Eq{CIRC4}.
The input events are vectors ${\bf y}_{n+1}=(\cos 30^\circ,\sin 30^\circ)$.
The direction of the initial vector ${\bf x}_{0}$ is chosen at random.
In this simulation $\alpha=0.99$.
For $n>60$ the ratio of the number of increments ($\Theta_{n+1}=0$)
to decrements ($\Theta_{n+1}=1$) is 1/3, which is $(\sin 30^\circ/\cos 30^\circ)^2$.
Data for $n<20$ has been omitted to show the oscillating behavior more clearly.
Lines are guides to the eyes.
Right:
The number of ($\Theta_{n+1}=1$) events
divided by the total number of events
as a function of the value of the input variable $\phi$.
Bullets: Each data point is obtained from a
simulation of 1000 events with a fixed, randomly
chosen value of $0\le\phi<360^\circ$, using the last 500 events
to count the number of ($\Theta_{n+1}=1$) events.
Solid line: $\cos^2\phi$.
}
\label{c30}
\end{center}
\end{figure*}

In general the behavior of the \DLM\ defined by rules
Eqs.~\Eq{CIRC3} and \Eq{CIRC4} is difficult to analyze without the use of a computer.
However, for a fixed input vector ${\bf y}_{n+1}={\bf y}$
it is clear what the \DLM\ will try to do:
It will minimize the cost Eq.~\Eq{CIRC4} by rotating
its internal vector ${\bf x}_{n+1}$ to bring it as close as possible to ${\bf y}$.
However, ${\bf x}_{n+1}$ will not converge to a limiting value but instead
it will keep oscillating about the input value ${\bf y}$.
An example of a simulation is given in Fig.~\ref{c30} (left).
For a fixed input vector ${\bf y}_{n+1}={\bf y}$
the \DLM\ reaches a stationary state in which its internal vector oscillates about ${\bf y}$.
In this stationary state the output signal consists
of a finite sequence of ones and zeros.
The \DLM\ repeats this sequence over and over again.
Obviously, the whole process is deterministic.
The details of the approach to the stationary state depend on
the initial value of the internal vector ${\bf x}_0$,
but the stationary state itself does not.

These observations are of much more general nature
than the example given in Fig.~\ref{c30} (left) suggests.
In fact, as the applications discussed below amply illustrate,
the stationary-state analysis is a very
useful tool to predict the behavior of the \DLMS.
Assuming that $0<1-\alpha\ll1$ and that we have reached
the stationary regime in which the internal vector performs
small oscillations about $(\cos\phi,\sin\phi)$,
a simple calculation shows that

\begin{eqnarray}
\delta\phi_0&=&\phi_{1,n+1}-\phi_{1,n}\approx\frac{1-\alpha^2}{2}\frac{\cos\phi}{\sin\phi}
\quad\hbox{if } \Theta_{n+1}=0,
\nonumber \\
\delta\phi_1&=&\phi_{1,n+1}-\phi_{1,n}\approx\frac{\alpha^2-1}{2}\frac{\sin\phi}{\cos\phi}
\quad\hbox{if } \Theta_{n+1}=1.
\label{CIRC5a}
\end{eqnarray}
In the stationary regime, we have $N_0\delta\phi_0\approx N_1\delta\phi_1$
where $N_0$ ($N_1$) is the number of $\Theta_{n+1}=0$
($\Theta_{n+1}=1$) events.
From Eq.~\Eq{CIRC5a} it then follows immediately
that $N_0/(N_0+N_1)\approx \sin^2\phi$ and $N_1/(N_0+N_1)\approx \cos^2\phi$.
The results of this analysis are in excellent agreement with
the simulation results shown in Fig.~\ref{c30} (right).

The conventional approach to regard the variables $\Theta_{n+1}$
as input is fundamentally different from the approach adopted in this paper.
This can be seen by reformulating the update rules in terms of difference equations
and to assume that the $\Theta_{n+1}=0,1$ are independent, uniform random variables
with mean $\Theta=\EXPECT{\Theta_{n+1}}$.
The four rules Eq.~\Eq{CIRC3} can be written as

\begin{eqnarray}
x_{1,n+1}^2&=&\alpha^2 x_{1,n}^2 + (1-\alpha^2)\Theta_{n+1},
\nonumber \\
x_{2,n+1}^2&=&\alpha^2 x_{2,n}^2 + (1-\alpha^2)(1-\Theta_{n+1}).
\label{CIRC5}
\end{eqnarray}
Formally Eq.~\Eq{CIRC5} has the same structure as Eq.~\Eq{ILLU4}.
Averaging over many realizations of $\{\Theta_{n+1}=0,1\}$
and taking the limit $n\rightarrow\infty$ we obtain

\begin{eqnarray}
\EXPECT{x_{1}^2}&=&\lim_{n\rightarrow\infty}\EXPECT{x_{1,n+1}^2}=\Theta,
\nonumber \\
\EXPECT{x_{2}^2}&=&\lim_{n\rightarrow\infty}\EXPECT{x_{2,n+1}^2}=1-\Theta.
\label{CIRC6}
\end{eqnarray}
In other words, a machine that operates according to the rules Eq.~\Eq{CIRC3}
and receives as input the random sequence $\Theta_{n+1}$ will (on average)
approach a state in which the direction of its internal vector
gives us an estimate of the $\Theta=\EXPECT{\Theta_{n+1}=0,1}$.
In contrast, a \DLM\ that minimizes the cost Eq.~\Eq{CIRC4} and
updates its internal state according to Eq.~\Eq{CIRC3}
responds on either output channel
$\Theta_{n+1}=0$
or output channel
$\Theta_{n+1}=1$,
with a frequency that is directly related
to the difference between the current input angle and the angle defined by the internal vector.

\subsection{Learning points on a $K$-dimensional hypersphere}\label{HYP}

Consider a sequence of events, characterized by
vectors ${\bf y}_{n+1}=(y_{1,n+1},y_{2,n+1},\ldots,y_{K,n+1})$ for $n>0$.
The vector ${\bf y}_{n+1}$ is the input for the \DLM.
The internal state of the \DLM\ is described
by a $K$-dimensional unit vector ${\bf x}_n=(x_{1,n},x_{2,n},\ldots,x_{K,n})$.
We define the $2K$ candidate update rules
$\{ j=1,\ldots,K; s_j=\pm1\}$ by

\begin{eqnarray}
x_{i,n+1}&=&s_j\sqrt{1+\alpha^2(x_{i,n}^2-1)}
\quad\hbox{if}\quad i=j,
\nonumber \\
x_{i,n+1}&=&\alpha x_{i,n}
\quad\hbox{if}\quad i\not=j.
\label{HYP1}
\end{eqnarray}
Note that ${\bf x}^T_{n}{\bf x}^{\phantom{T}}_{n}=1$ implies
${\bf x}^T_{n+1}{\bf x}^{\phantom{T}}_{n+1}=1$ for each of the $2K$ update rules.
The \DLM\ responds to the input ${\bf y}_{n+1}$ by
selecting from the $2K$ possible rules in Eq.~\Eq{HYP1},
the update rule that minimizes the cost

\begin{equation}
C = -{\bf x}^T_{n+1}{\bf y}^{\phantom{T}}_{n+1}
,
\label{HYP2}
\end{equation}
and by sending a message
containing ${\bf y}_{n+1}$ (or, depending on the application, ${\bf x}_{n+1}$)
on one of its output channels.
Note that the minimum of the cost function Eq.~\Eq{HYP2} does not depend on
the length of the vectors ${\bf x}_{n+1}$ or ${\bf y}_{n+1}$.
Disregarding the variables $s_j$ that merely serve to determine the sign
of $x_{i,n+1}$ there are $K$ rules. Hence there can be as many as $K$ output channels.
However, depending on the application, it may be expedient to reduce the number
of output channels by arranging them in groups.

\subsection{Communication between events}\label{TWOONE}

The \DLMS\ analyzed in the previous subsections have one input channel that receives input
and two output channels, only one of which sends out data (a message) at a particular event.
An obvious generalization is to construct \DLMS\ that accept, at a given instance,
input from one out of two different sources.
This is absolutely necessary if we want to build machines in which events can communicate
or, in physical terms, interact with each other.
We now demonstrate that the \DLMS\ that we introduced above
already have the capability to let events interact with each other.
Therefore we do not need to add a new feature or rule to the \DLMS.

Consider a \DLM\ that has two input channels 0 and 1
and an internal vector ${\bf x}_{n}$ with $K=4$ components.
At the $n+1$-th event, either input channel 0 receives the two-component vector
${\bf y}_{n+1}=(y_{1,n+1},y_{2,n+1})$
or input channel 1 receives the two-component vector
${\bf y}_{n+1}=(y_{3,n+1},y_{4,n+1})$.

In the former case the \DLM\ transforms this input into the input vector
${\bf \hat y}_{n+1}=(y_{1,n+1},y_{2,n+1},x_{3,n},x_{4,n})$
of four elements by using the current internal vector as a source for the missing elements.
Similarly, in the latter case the input vector becomes
${\bf \hat y}_{n+1}=(x_{1,n},x_{2,n},y_{3,n+1},y_{4,n+1})$.
Then the \DLM\ uses ${\bf \hat y}_{n+1}$ to determine
the cost and selects the update rule according to the
procedure described in Section~\ref{HYP} (with ${\bf \hat y}_{n+1}$ replacing ${\bf y}_{n+1}$).
This \DLM\ learns the two-dimensional vectors
${\bf y}_{n+1}=(y_{1,n+1},y_{2,n+1})$
and
${\bf y}_{n+1}=(y_{3,n+1},y_{4,n+1})$
separately, as if it consists of two separate, independent
two-dimensional \DLMS, with the additional crucial feature
that the internal vector represents a point on a 4-dimensional unit sphere.

It is not difficult to imagine what this \DLM\ does in the case that
it receives events on only one of the two input channels (say 0).
Irrespective of the initial value of the internal vector
${\bf x}_{0}$, the \DLM\ will always select the update rule with $j=1,2$
(see Eq.~\Eq{HYP1})
and the two components $x_{3,n}$ and $x_{4,n}$ will vanish exponentially
fast with increasing $n$ (recall that $0<\alpha<1$).
Thus, after a few events the internal state of the \DLM\ indicates
that the \DLM\ receives events on only one channel.

If the machine receives input on both channels (but never simultaneously),
Eq.~\Eq{HYP1} implies that the \DLM\ only scales
the two components of the internal state
that it uses to provide the missing elements
for building the input ${\bf \hat y}_{n+1}$.
Therefore, in the stationary regime, the
length of the two-dimensional vector $(x_{1,n},x_{2,n})$ ($(x_{3,n},x_{4,n})$)
is proportional to the number of events on input channel 0 (1).
Furthermore the number of $j=1,2$ ($j=3,4$) events is approximately
equal to the number of events on input channel 0 (1).
Although this may seem a very elementary form
of communication, it is sufficient to construct
\DLMS\ that perform very complicated tasks.

\subsection{Summary}
The \DLMS\ described above are simple deterministic machines that make decisions.
The \DLM\ responds to the input event by
choosing from all possible alternatives, the internal state
that minimizes the error between the input and the internal state itself.
Then the \DLM\ sends a message through one of its output channels.
The message contains information about the decision the \DLM\ took
while updating its internal state and, depending on the application,
also contains other data that the \DLM\ can provide.
By updating its internal state, the \DLM\ ``learns" about the input it receives
and by sending messages through one of its two output channels, it tells
its environment about what it has learned.
In the sequel we will call such a machine a
{\bf deterministic learning machine} (DLM).
For a particular choice of the update rule (see Section~\ref{ILLUa}),
the \DLM\ performs linear estimation but as the other examples
of this Section amply demonstrate, minor modifications to this rule
and/or cost function yield \DLMS\ that may behave
in a substantially different manner.

\def\DLM{DLM}
\def\DLMS{DLMs}
\section{Application to blind classification}\label{NDIM}

The \DLM\ of Section~\ref{ILLUa} learns about the input data by moving a point on a line.
Obviously, this point separates two parts of the line.
The generalization to $K$-dimensional space is a
$(K-1)$-dimensional hyperplane that divides the space into two parts.
Thus, to interpret two-dimensional data the \DLM\
should learn a line instead of a point.
We represent the line by a segment $L_n$ defined by
its mid-point ${\bf x}_n$ and its direction ${\bf d}_n$.
As the \DLM\ receives an event ${\bf y}_{n+1}$, i.e. a point in a two-dimensional plane,
the \DLM\ updates its internal line segment $L_n$
and sends the information describing $L_n$
through the -1 (+1) channel, depending on whether
it lies on the left (right) side of the line.
The update procedure consists of two steps.
First we define two support points ${\bf v}_1$ and ${\bf v}_2$
on either side of ${\bf x}_n$ along the direction ${\bf d}_n$ by

\begin{eqnarray}
{\bf v}_1 &=& {\bf x}_n - {\bf d}_n/2,
\nonumber \\
{\bf v}_2 &=& {\bf x}_n + {\bf d}_n/2,
\label{NDIM1}
\end{eqnarray}
and we update the two support points according to

\begin{eqnarray}
{\bf \hat{v}}_1 &=& {\bf v}_1 + (1-\alpha)({\bf y}_{n+1}-{\bf v}_1)\Vert {\bf y}_{n+1}-{\bf v}_1\Vert ,
\nonumber \\
{\bf \hat{v}}_2 &=& {\bf v}_2 + (1-\alpha)({\bf y}_{n+1}-{\bf v}_2)\Vert {\bf y}_{n+1}-{\bf v}_2\Vert ,
\label{NDIM2}
\end{eqnarray}
where $0<\alpha<1$ controls the learning process.
Then we compute the new mid-point and direction of the line segment:

\begin{eqnarray}
{\bf x}_{n+1} &=& ({\bf \hat{v}}_1+{\bf \hat{v}}_2)/2,
\nonumber \\
{\bf d}_{n+1} &=& ({\bf \hat{v}}_1-{\bf \hat{v}}_2)/\Vert {\bf \hat{v}}_1-{\bf \hat{v}}_2 \Vert.
\label{NDIM3}
\end{eqnarray}
From Eq.~\Eq{NDIM2} it follows that the support point farthest away from ${\bf y}_{n+1}$ makes
the largest move.
Therefore, as new input data is received by the \DLM,
both the mid-point and the direction of the line segment change.
Note that the update rule Eq.~\Eq{NDIM2} is non-linear
in the difference between internal and input vector.
Although a linear update rule also works,
our numerical experiments (results not shown)
indicate that the non-linear rule Eq.~\Eq{NDIM2}
performs much better.

In general ${\bf x}_n$ will converge to the mean of the input vectors and
${\bf v}_1$ and ${\bf v}_2$ will be pulled most strongly in the
direction of largest variance.
Therefore $L_n$ will be (approximately) perpendicular
to the largest principal component of the covariance matrix of the input data.
In other words, the \DLM\ defined above
can find the eigenvector that corresponds to the largest
eigenvalue of the covariance matrix
by processing data points in a sequential manner, i.e., without actually
having to compute the elements of the covariance matrix.

\begin{figure}[t]
\begin{center}
\setlength{\unitlength}{1cm}
\begin{picture}(14,10)
\put(0,5){\includegraphics[width=7cm]{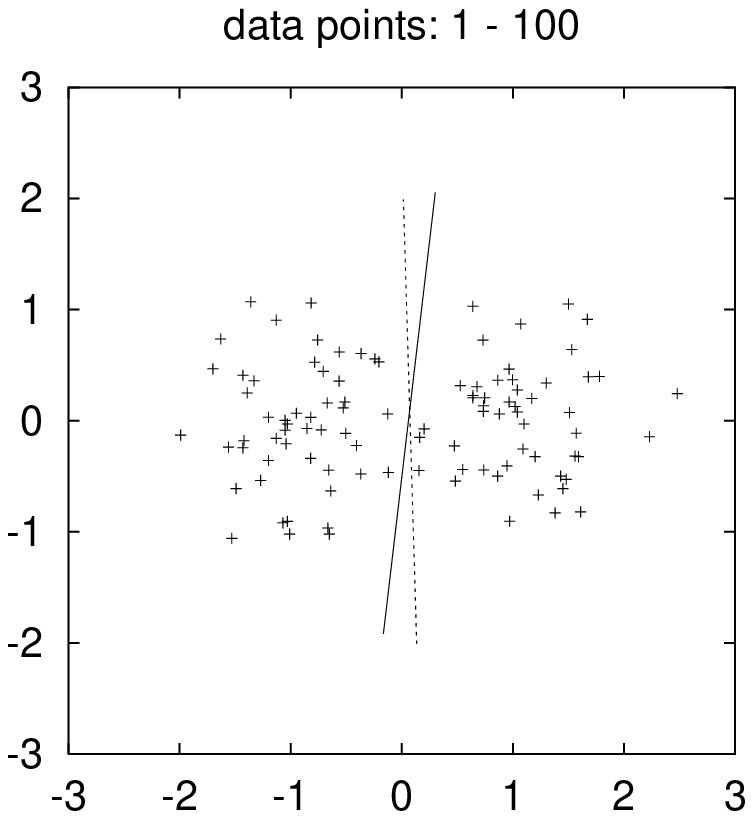}}
\put(5,5){\includegraphics[width=7cm]{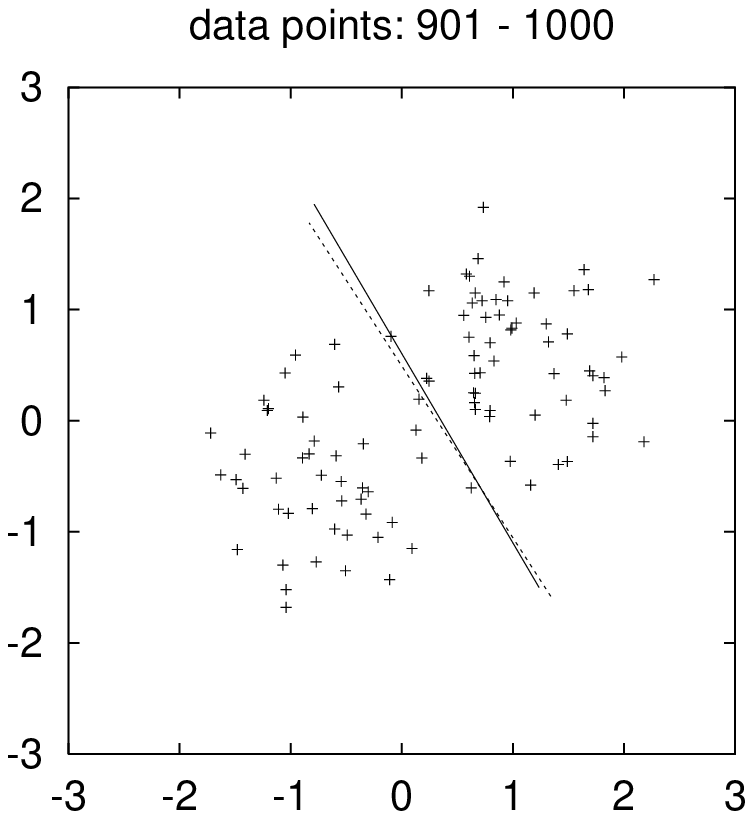}}
\put(10,5){\includegraphics[width=7cm]{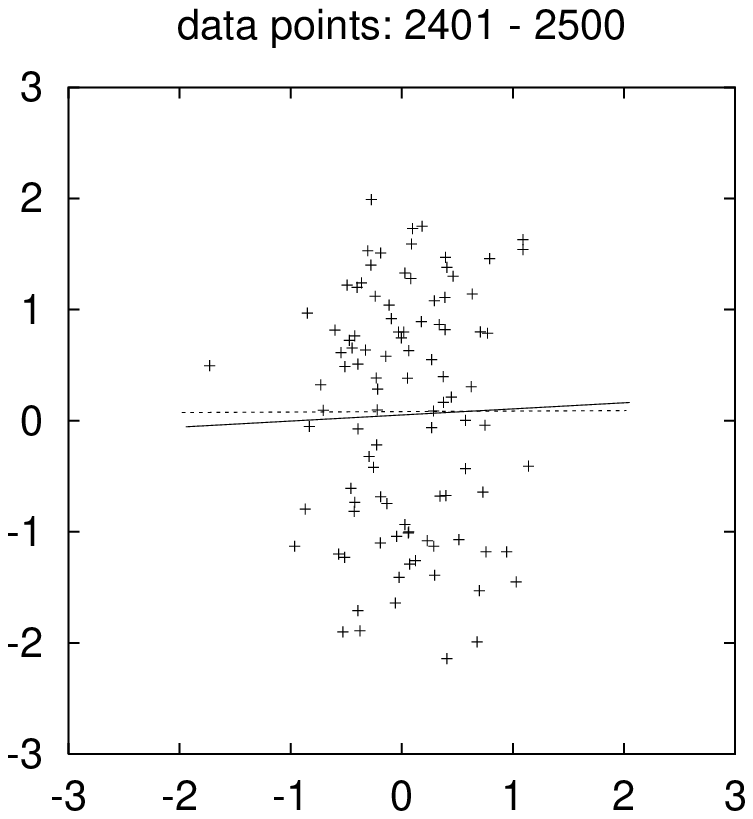}}
\put(0,0){\includegraphics[width=7cm]{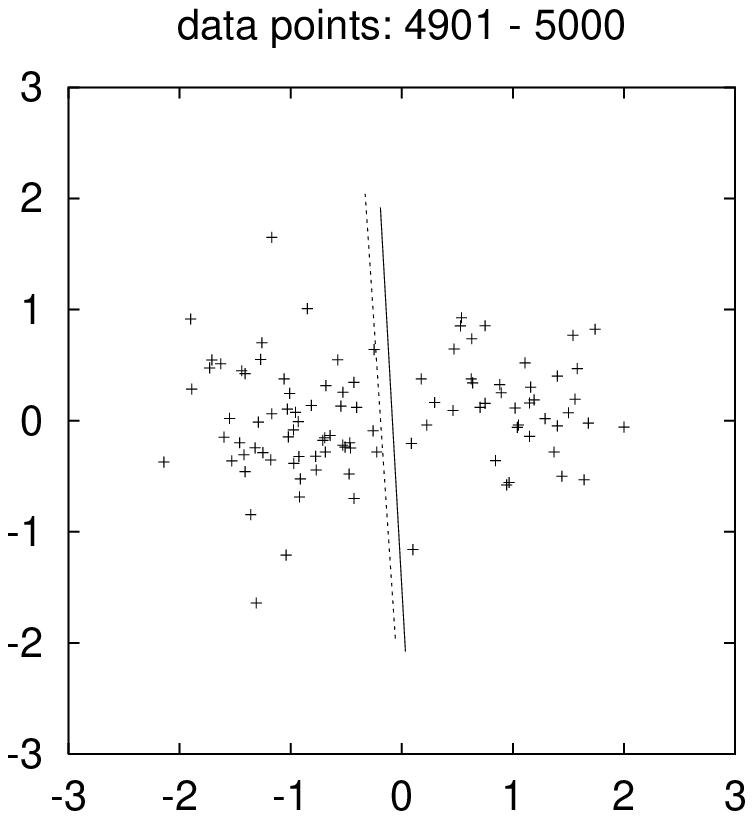}}
\put(5,0){\includegraphics[width=7cm]{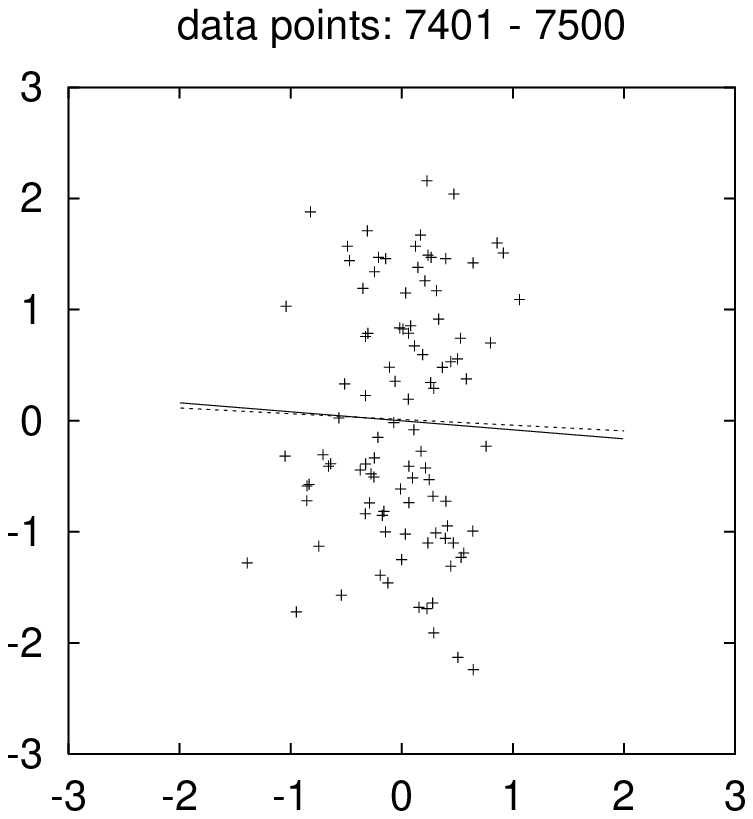}}
\put(10,0){\includegraphics[width=7cm]{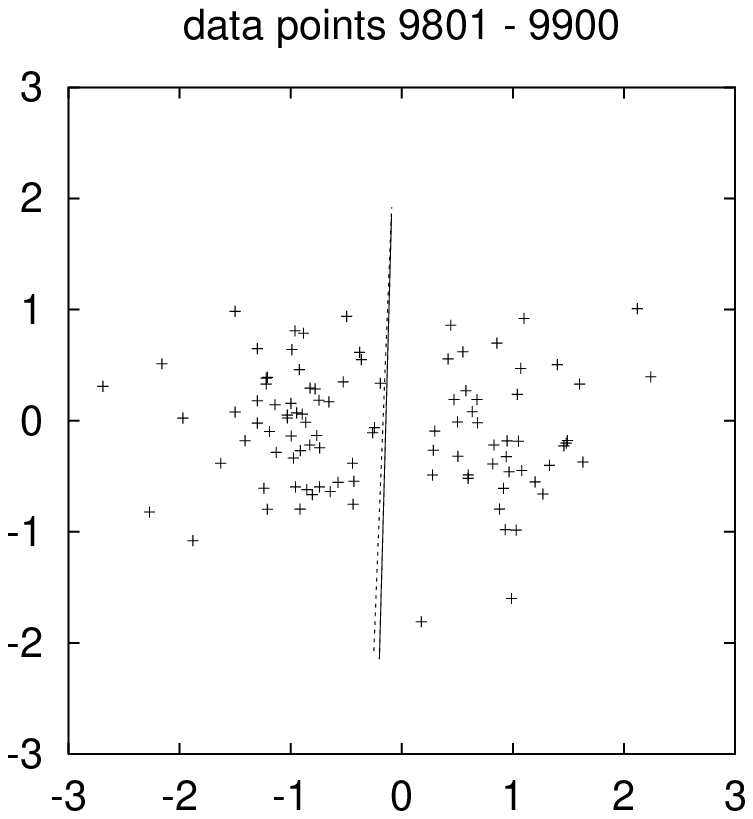}}
\end{picture}
\caption{Snapshots of the input data and results
of a \DLM-based classifier defined by Eqs.~\Eq{NDIM1} -- \Eq{NDIM3}
(solid line) and a conventional
principal-component-based classifier (dashed line)~\cite{MARD82}.
The data points are random deviates with a normal
distribution with variance $1/2$ and
means $\pm(\cos(2\pi n/10000),\sin(2\pi n/10000))$.
Each panel shows the output of the \DLM-based classifier
after it has processed, point-by-point, the 100 data points shown.
The classifier smoothly follows the rotation of the means.
In contrast to the event-by-event processing
of the \DLM-based classifier,
the principal-component-based classifier processes
the whole set of 100 data points simultaneously.
}
\label{2dp2}
\end{center}
\end{figure}

As an illustration of the capabilities of the
\DLM\ introduced in this section,
let us consider a classification task in which we want to blindly group events
into two categories.
The input data
${\bf y}_{n+1}=(y_{1,n+1},y_{2,n+1})$
are generated through a Gaussian random process described by:

\begin{eqnarray}
y_{1,n} &=& \cos (\gamma n+s)\pi + r_1,
\nonumber \\
y_{2,n} &=& \sin (\gamma n+s)\pi + r_2,
\label{NDIM4}
\end{eqnarray}
where $s$ is a uniform random bit.
The random numbers $r_1$ and $r_2$ are drawn from
the normal distribution $N(0,1/2)$.
In our numerical example we take $\gamma = 1/5000$ and $\alpha=0.99$.
From Eq.~\Eq{NDIM4} it is clear that the
input events consist of points in a plane that are
drawn from one of two ($s=0,1$) Gaussian distributions,
the centers of which rotate with a period of 10000 events.
The mean of all input data is $(0,0)$ and
there is no preferred direction of largest variance.
The reason of course is that the center of the Gaussian distributions
slowly moves on the unit circle.
Clearly, this kind of classification task can only be performed by
permanently updating the estimate of the direction
and that is exactly what the \DLM\ does.
In Fig.~\ref{2dp2} we present results of a
blind classification experiment that illustrates the operation
of the \DLM\ defined by the rules Eqs.~\Eq{NDIM1} -- \Eq{NDIM3}.
The \DLM\ processes event-by-event, each time updating its
estimate for the separatrix.
For comparison we also show the result obtained
by the principal component analysis~\cite{MARD82} using as input
the group of 100 most recent data points processed by the \DLM.
The differences between both classifiers are rather small so that
it is clear that the \DLM-based classifier performs very well.

The two-dimensional \DLM\ described above can easily be extended to
a \DLM\ that processes $K$-dimensional input data.
Instead of a line segment the \DLM\ has to learn a
segment of a $(K-1)$-dimensional hyperplane.
This can be done by extending the procedure used in the two-dimensional case.
The hyperplane segment is described by
a mid-point ${\bf x}_n$ and $K-1$ orthonormal directions ${\bf d}_k$
for $k=1,\ldots,K-1$.
We choose $K$ points $\{{\bf v}_k\}$ on the hyperplane defined by
$\{{\bf d}_k\}$ and ${\bf x}_n$ such that the distance between each pair of points is one.
As new input data ${\bf y}_{n+1}$ is received by the \DLM\,
these points are updated according to (the generalization of) Eq.~\Eq{NDIM2}.
As in the two-dimensional case,
from the updated points we can calculate the new mid-point and the new
directions.
However, unlike in the two-dimensional case, these
directions do not need to be orthonormal.
The orthonormality is then restored by using the (modified)
Gramm-Schmidt procedure~\cite{GOLU96}.

\begin{figure*}[t]
\begin{center}
\setlength{\unitlength}{1cm}
\begin{picture}(14,7)
\put(-1.75,0){\includegraphics[width=8cm]{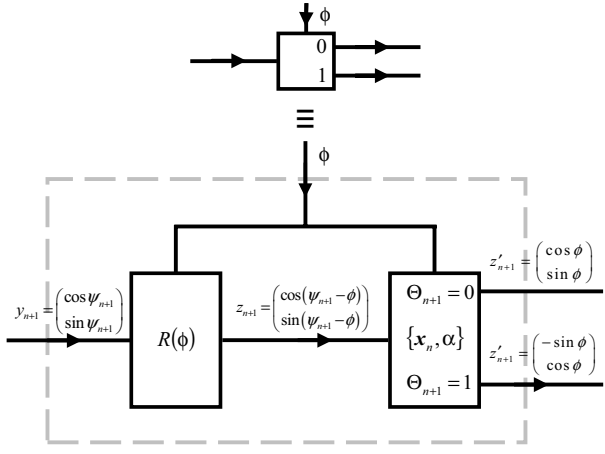}}
\put(7,0){\includegraphics[width=9cm]{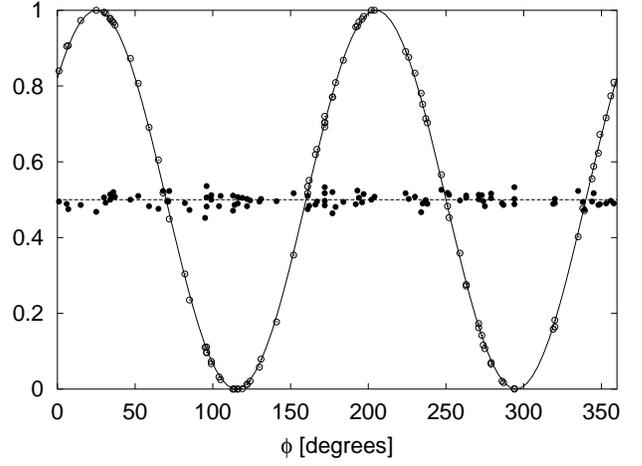}}
\end{picture}
\caption{
Left:
Diagram of the \DLM\ network that simulates a polarizer
on a deterministic, event-by-event basis.
Right:
Simulation results for the \DLM\ network shown on the left. 
Each data point represents the number of events
in an output channel accumulated after 1000 input events.
After each set of 1000 events, the orientation $\phi$ of the polarizer
is changed randomly.
Open circles: Normalized intensity in output channel 0
for incoming photons with a polarization angle $\psi=25^\circ$;
Solid line: Result ($\cos^2(\psi-\phi)$) obtained from quantum theory~\cite{QuantumTheory}
for incoming photons with a polarization angle $\psi=25^\circ$;
Bullets: Normalized intensity in output channel 1
for incoming photons with a random polarization angle $\psi$;
Dashed line: Result of quantum theory~\cite{QuantumTheory}
for incoming photons with a random polarization angle $\psi$.
}
\label{figpol}
\label{p25}
\end{center}
\end{figure*}

\setlength{\unitlength}{1cm}
\begin{figure*}[t]
\begin{center}
\setlength{\unitlength}{1cm}
\begin{picture}(14,7)
\put(-1.,0){
\includegraphics[width=6cm]{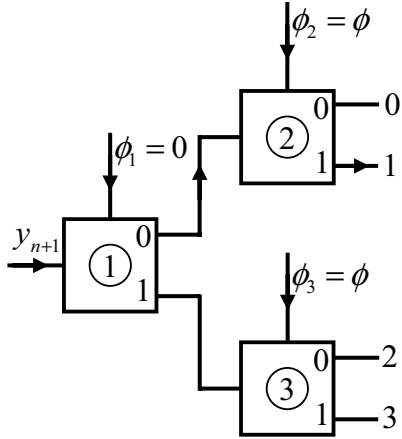}
}
\put(7,0){
\includegraphics[width=9cm]{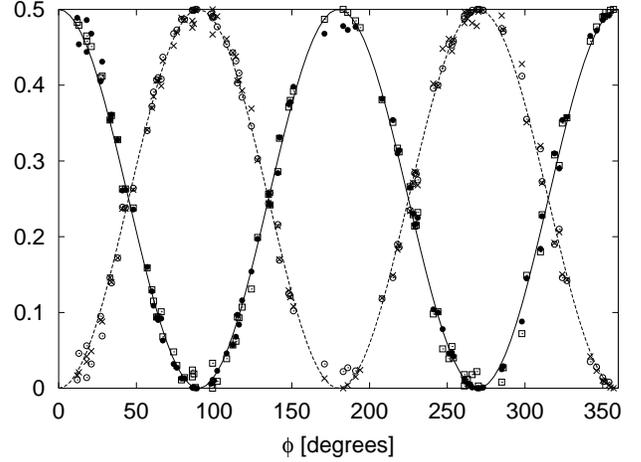}
}
\end{picture}
\caption{
Left:
Schematic representation of an experiment with three polarizers~\cite{FEYN82}.
Right:
Simulation results for the network of \DLMS\ shown on the left. 
Each data point represents the normalized intensity accumulated over 1000 events.
After each set of 1000 events, the orientation $\phi$ of the polarizers 2 and 3
is changed randomly.
Bullets: Output channel 0;
Crosses: Output channel 1;
Open circles: Output channel 2;
Open squares: Output channel 3.
Lines represent the results of quantum theory~\cite{QuantumTheory}.
}
\label{fig3pol}
\label{3pol}
\end{center}
\end{figure*}

\section{Application to deterministic simulation of quantum interference}\label{QI}
\subsection{Photon polarization}\label{POLA}

We demonstrate that the \DLM\ defined by Eqs.~\Eq{CIRC3} and \Eq{CIRC4}
and a passive element that performs a plane rotation
are sufficient to perform a deterministic simulation
of the quantum theory~\cite{QuantumTheory} of photon polarization.

We start by recalling some elementary facts about photon polarization~\cite{BAYM74,FEYN82}.
Some optically active materials like calcite
split an incoming beam of light into two spatially separated beams~\cite{BORN64,FEYN82}.
The light intensity of these beams is related to the angle of polarization
$\psi$ of the electromagnetic wave, relative to the orientation $\phi$
of the material~\cite{BORN64}.
We disregard all imperfections of real experiments
and assume that the experimental data are in exact agreement with the wave mechanical theory.
Then the intensities $I_0$ of beam 0 and $I_1$ of beam 1 are given by~\cite{BAYM74,FEYN82}

\begin{eqnarray}
I_0&=&\cos^2(\psi-\phi)\quad,\quad I_1=\sin^2(\psi-\phi),
\label{POLA1}
\end{eqnarray}
respectively.
If the incident beam has a random polarization, averaging of Eq.~\Eq{POLA1}
over all $\psi$ shows that half of the light intensity will go to beam 0
and the other half to beam 1.

If the conventional light source is replaced by a source that
emits one photon at a time, the photon leaves the material
either in the direction of beam 0 or beam 1, never in both~\cite{FEYN82}.
Collecting photons over a sufficiently long period shows that
Eq.~\Eq{POLA1} still gives the number of photons detected in the direction of beam 0 (1),
divided by the total amount of detected photons~\cite{FEYN82}.
Quantum theory~\cite{QuantumTheory} describes the polarization in terms of a
two-dimensional (complex-valued) vector
and the action of the material is to rotate this vector
by an angle $\phi$ (set by the experimentalist)~\cite{BAYM74}.
The probability to observe photons in beam 0 (1) is given
by the square of the 0-th (1-st) element of the vector~\cite{BAYM74}.
In addition, as the photon leaves the material in beam 0 (1), its
polarization is $\phi$ ($\phi+\pi/2$)~\cite{BAYM74}.
Thus the piece of material can be used to prepare and also determine the polarization
of the photons and is called a ``polarizer''~\cite{BORN64}.

According to quantum theory~\cite{QuantumTheory}, the polarizer rotates the vector of polarization amplitudes
in the following manner~\cite{BAYM74}:

\begin{eqnarray}
\left(
\begin{array}{c}
b_0\\b_1
\end{array}
\right)
&=&
\left(
\begin{array}{cc}
\cos\phi&\sin\phi\\
-\sin\phi&\cos\phi
\end{array}
\right)
\left(
\begin{array}{c}
a_0\\a_1
\end{array}
\right).
\label{POLA2}
\end{eqnarray}
Still according to quantum theory~\cite{QuantumTheory}, the intensity in beam 0 (1) is given by
$|b_0|^2$ ($|b_1|^2$).
An incident beam with an angle of polarization $\psi$ is described
by the vector $(a_0,a_1)=(\cos\psi,\sin\psi)$.
From Eq.~\Eq{POLA2} we obtain $(b_0,b_1)=(\cos(\psi-\phi),\sin(\psi-\phi))$
and hence $I_0=|b_0|^2=\cos^2(\psi-\phi)$ and
$I_1=|b_1|^2=\sin^2(\psi-\phi)$, in agreement with Eq.~\Eq{POLA1}.

We now construct a simple deterministic machine that generates
events of which the distribution agrees with the probability distributions
predicted by quantum theory~\cite{QuantumTheory}.
The layout of this ``polarizer'' is shown in Fig.~\ref{figpol}.
The incoming event (photon) carries an (unknown) angle $\psi_{n+1}$.
The purpose of the passive element $R(\phi)$ is to perform a
rotation

\begin{eqnarray}
R(\phi)&=&
\left(
\begin{array}{cc}
\cos\phi&-\sin\phi\\
\sin\phi&\cos\phi
\end{array}
\right),
\label{Rphi}
\end{eqnarray}
of the input vector ${\bf y}_{n+1}=(\cos\psi_{n+1},\sin\psi_{n+1})$ by the angle $\phi$.
The resulting vector ${\bf z}_{n+1}=(\cos(\psi_{n+1}-\phi),\sin(\psi_{n+1}-\phi))$
is sent to the input of a \DLM\ that operates according to
Eqs.~\Eq{CIRC3} and \Eq{CIRC4}.
If $\Theta_{n+1}=0$, the \DLM\ responds by sending
the vector ${\bf z'}_{n+1}=(\cos\phi,\sin\phi)$
through the output channel 0.
If $\Theta_{n+1}=1$, the \DLM\ responds by sending
the vector ${\bf z'}_{n+1}=(\cos(\phi+\pi/2),\sin(\phi+\pi/2))$
through the output channel 1.
Clearly this procedure is strictly deterministic.
We emphasize that the \DLM\ processes information event by event
and does not store the data contained in each event.

In Fig.~\ref{p25} (right)
we show simulation results for the machine depicted in Fig.~\ref{figpol} (left).
Each data point represents the intensity in beam 0 (1), i.e., the
number of $\Theta=0$ $(1)$ events divided by the total amount of events.
The machine is initialized once by choosing a random direction
of the vector ${\bf x}_{0}$.
The angle of rotation $\phi$ is kept fixed for 1000 events, then a
uniform random number is used to select another direction, and
this procedure is repeated 100 times.
In all these numerical experiments we set $\alpha=0.99$.
Fig.~\ref{p25} shows the results for two different numerical experiments:
In the first set of 100 runs, the direction of polarization $\psi$ of
the incoming photons is also determined by means of uniform random numbers.
In the second set of 100 runs, the direction of polarization of
the incoming photons is fixed ($\psi=25^\circ$).
From Fig.~\ref{p25} (right) it is clear that quantum theory~\cite{QuantumTheory} provides
a very good description of the input-output behavior
of the \DLM\ shown in Fig.~\ref{figpol} (left).

As a second illustration we use the same \DLM\ to simulate an
experiment with three polarizers described by Feynman~\cite{FEYN82}.
The diagram of this experiment is shown in Fig.~\ref{fig3pol}.
A randomly polarized beam of photons passes through the
first polarizer (without loss of generality we set its angle
$\phi_1$ equal to zero).
Each output channel is used as input to another polarizer.
Both these polarizers are tilted by the same
angle $\phi_2=\phi_3=\phi$.
According to quantum theory~\cite{QuantumTheory}, the intensity at the output
of these four channels is (from top to bottom, see Fig.~\ref{fig3pol})
$2^{-1}\cos^2\phi$,
$2^{-1}\sin^2\phi$,
$2^{-1}\sin^2\phi$,
and
$2^{-1}\cos^2\phi$.
The results of our numerical experiments are shown in Fig.~\ref{3pol}.
The simulation procedure is the same as the one used to
generate the data of Fig.~\ref{p25}.
Also in these numerical experiments we set $\alpha=0.99$.
We emphasize once more that the randomness in these
discrete-event simulations only enters through the
characterization of the photon source and through our
procedure of selecting the direction of the polarizer
for each set of 1000 events.
Actually, the latter only serves to counter the possible
objection that the apparent quantum mechanical behavior
would be caused by monotonically changing
the direction of the polarizers.
As in the previous example, it is clear that quantum theory~\cite{QuantumTheory} describes
the input-output behavior of the three-\DLM\ network very well.

\subsection{Beam splitter}\label{BS}

We now show that two $K=4$ \DLMS\ and two passive devices that perform a plane rotation
by $45^\circ$ are sufficient to build a network that behaves as if
it where a single-photon beam splitter.
First we describe the network and then we demonstrate that it acts as a beam splitter.

\setlength{\unitlength}{1cm}
\begin{figure*}[t]
\begin{center}
\setlength{\unitlength}{1cm}
\begin{picture}(14,7)
\put(-1.75,0){\includegraphics[width=8cm]{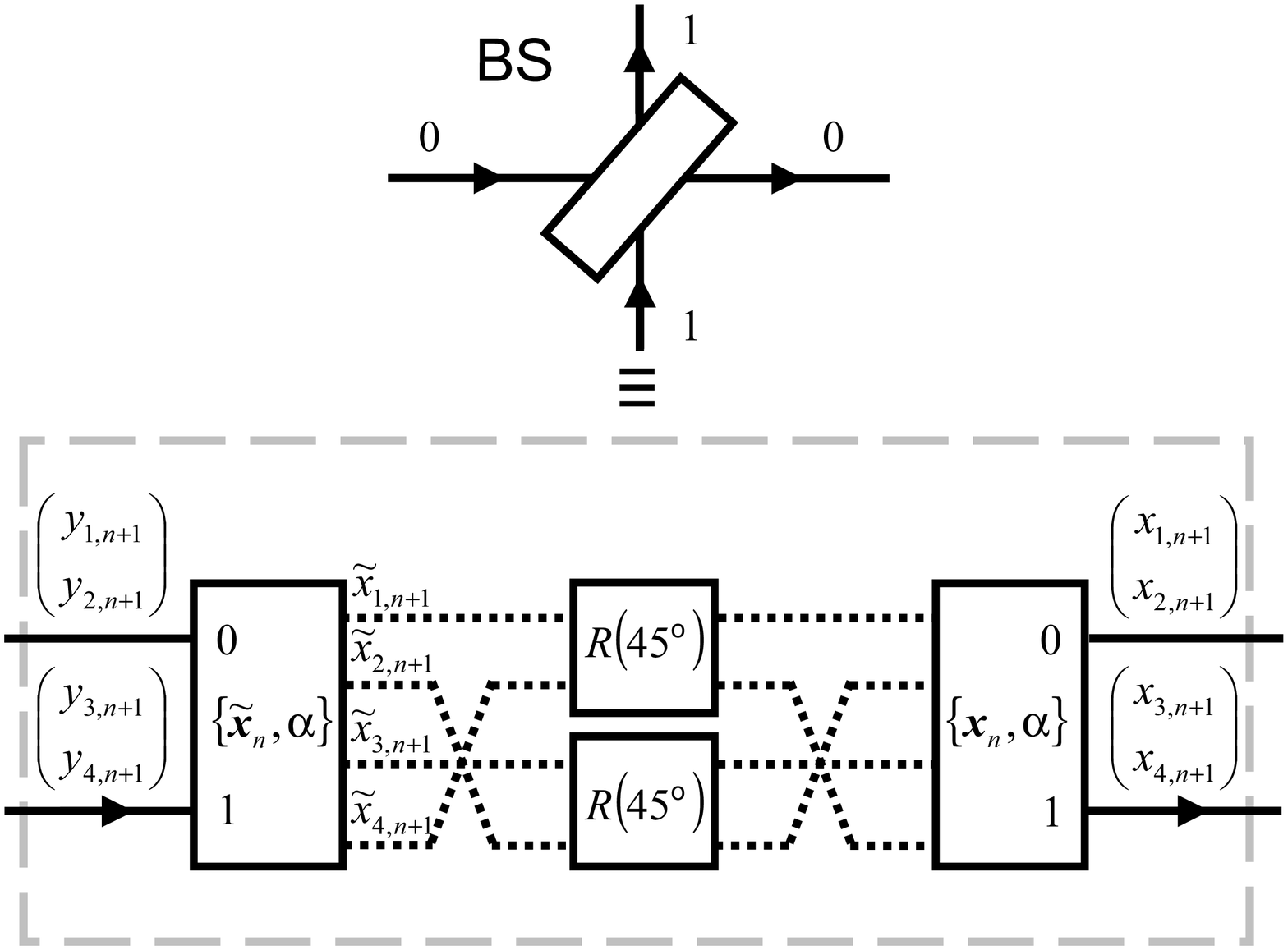}}
\put(7,0){\includegraphics[width=9cm]{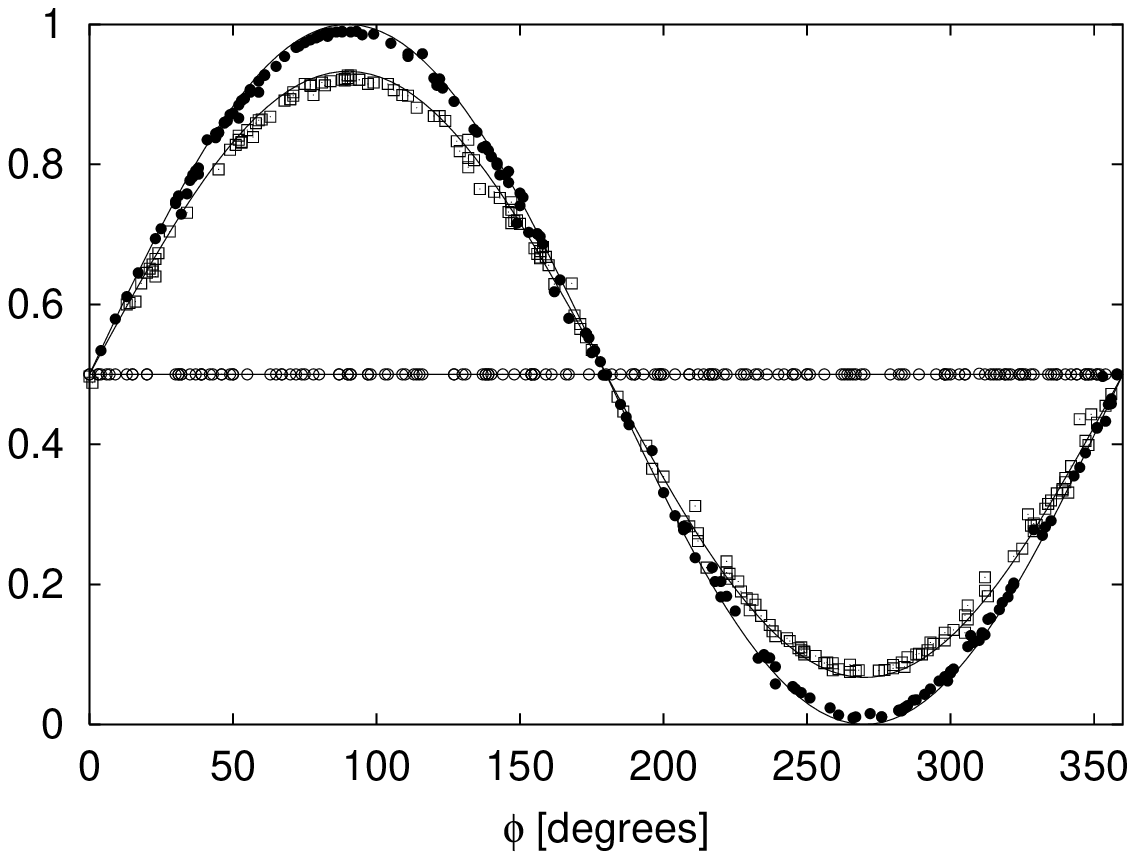}}
\end{picture}
\caption{
Left:
Diagram of the network of two \DLMS\ that performs a deterministic
simulation of a single-photon beam splitter (BS) on an event-by-event basis~\cite{MZIdemo}.
The solid lines represent the input and output channels of the BS.
Dashed lines indicate the flow of data within the BS.
Right:
Simulation results for the beam splitter shown on the left.
Input channel 0 receives $(y_{1,n+1},y_{2,n+1})=(\cos\psi_0,\sin\psi_0)$
with probability $p_0$.
Input channel 1 receives $(y_{3,n+1},y_{4,n+1})=(\cos\psi_1,\sin\psi_1)$ with probability
$p_1=1-p_0$.
Each data point represents 10000 events.
After each set of 10000 events, a uniform
random number in the range $[0,360]$ is used to
choose the angles $\psi_0$ and $\psi_1$.
Markers give the simulation results for the
normalized intensity in output channel 0 as a function of
$\phi=\psi_0-\psi_1$.
Open circles: $p_0=1$;
Bullets: $p_0=0.5$;
Open squares: $p_0=0.25$.
Lines represent the results of quantum theory~\cite{QuantumTheory}.
}
\label{figbs}
\label{one-bs}
\end{center}
\end{figure*}

\setlength{\unitlength}{1cm}
\begin{figure*}[t]
\begin{center}
\setlength{\unitlength}{1cm}
\begin{picture}(14,7)
\put(-1.75,0){\includegraphics[width=8cm]{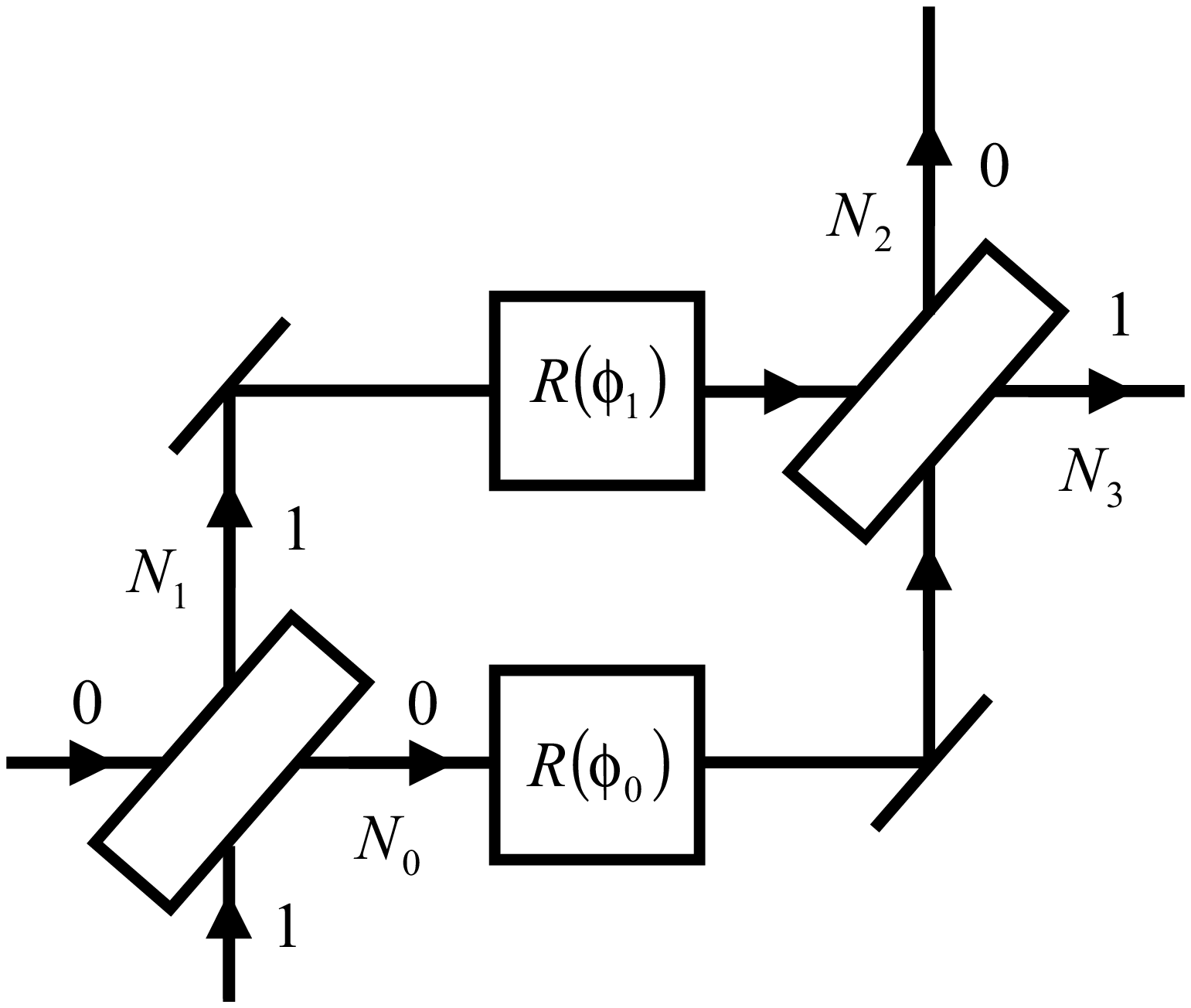}}
\put(7,0){\includegraphics[width=9cm]{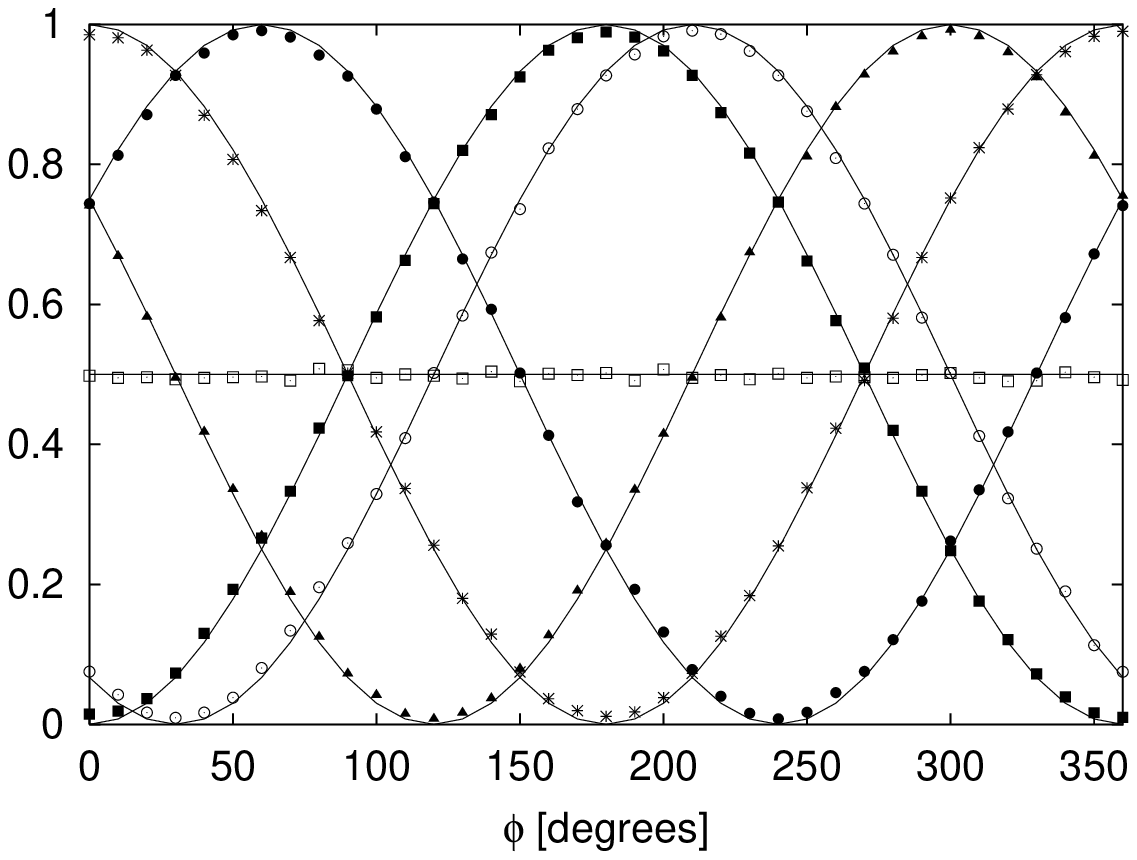}}
\end{picture}
\caption{
Left:
Diagram of a \DLM\ network that simulates a single-photon
Mach-Zehnder interferometer on an event-by-event basis~\cite{MZIdemo}.
The \DLM\ network consist of two BS devices (see Fig.~\ref{figbs} (left))
and two passive devices ($R(\phi_0)$ and $R(\phi_1)$)
that perform plane rotations by $\phi_0$ and $\phi_1$, respectively.
There is a one-to-one correspondence between
the elements of a physical Mach-Zehnder interferometer~\cite{BORN64,GRAN86}
and the units in the \DLM\ network.
The number of events $N_i$ in channel $i=0,\ldots,3$
corresponds to the probability for finding a photon
on the corresponding arm of the interferometer.
Right:
Simulation results for the \DLM-network shown on the left.
Input channel 0 receives $(y_{1,n+1},y_{2,n+1})=(\cos\psi_0,\sin\psi_0)$ with probability one.
A uniform random number in the range $[0,360]$ is used to choose the angle $\psi_0$.
Input channel 1 receives no events. 
Each data point represents 10000 events ($N_0+N_1=N_2+N_3=10000$).
Initially the rotation angle $\phi_0=0$ and after each set of 10000 events, $\phi_0$
is increased by $10^\circ$.
Markers give the simulation results for the normalized intensities
as a function of $\phi=\phi_0-\phi_1$.
Open squares: $N_0/(N_0+N_1)$;
Solid squares: $N_2/(N_2+N_3)$ for $\phi_1=0$;
Open circles: $N_2/(N_2+N_3)$ for $\phi_1=30^\circ$;
Bullets: $N_2/(N_2+N_3)$ for $\phi_1=240^\circ$;
Asterisks: $N_3/(N_2+N_3)$ for  $\phi_1=0$;
Solid triangles: $N_3/(N_2+N_3)$ for $\phi_1=300^\circ$.
Lines represent the results of quantum theory~\cite{QuantumTheory}.
}
\label{one-mz}
\label{figmz}
\end{center}
\end{figure*}

\setlength{\unitlength}{1cm}
\begin{figure*}[t]
\begin{center}
\includegraphics[width=14cm]{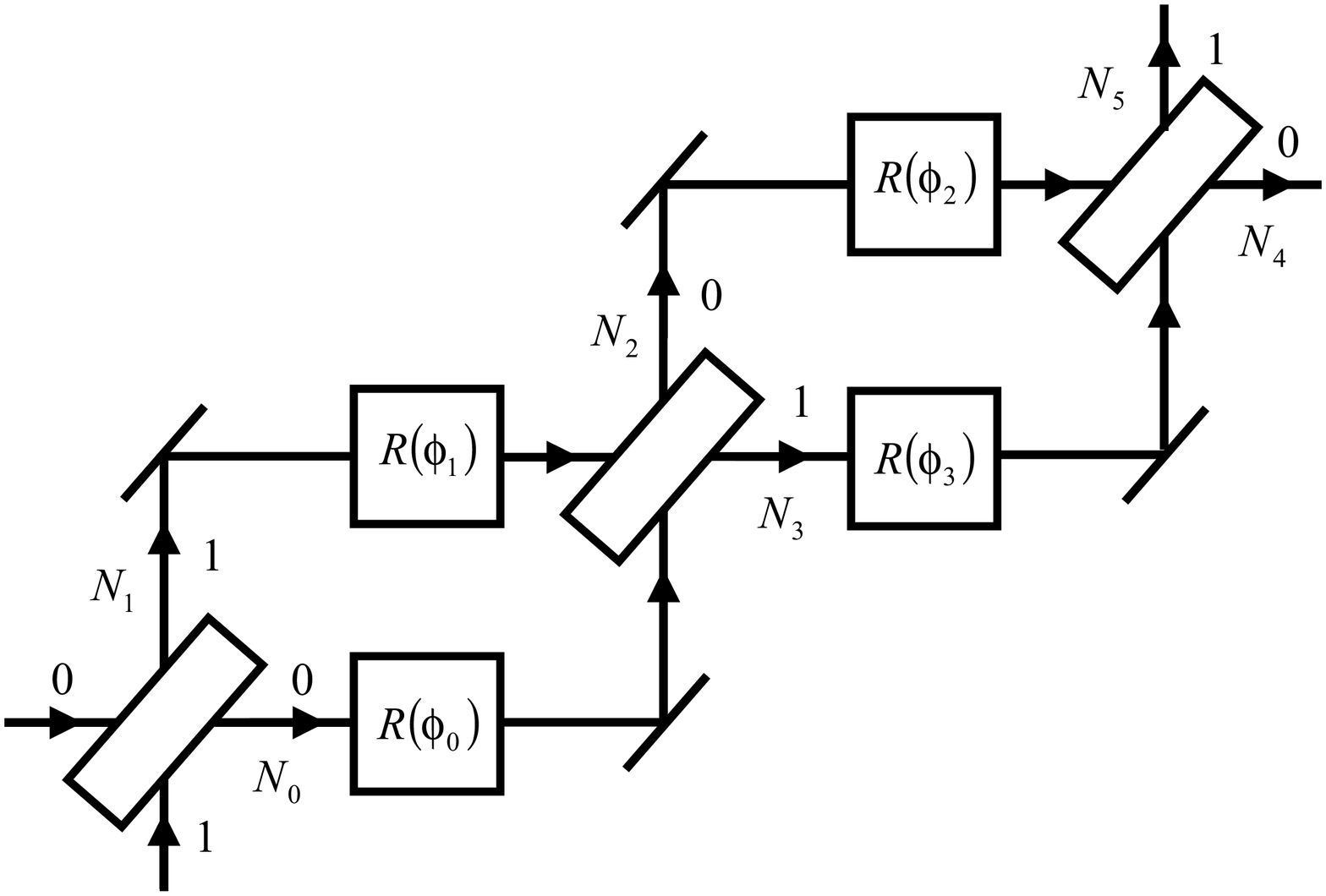}
\end{center}
\caption{Diagram of a \DLM\ network that simulates single-photon
propagation through two chained Mach-Zehnder interferometers
on an event-by-event basis.
}
\label{fig2mz}
\end{figure*}

The network shown in Fig.\ref{figbs} has two input channels (0 and 1) and
two output channels (0 and 1).
The network receives events at one of the two input channels.
Each input event carries information in the form of a two-dimensional unit vector.
Either input channel 0 receives $(y_{1,n+1},y_{2,n+1})$
or input channel 1 receives $(y_{3,n+1},y_{4,n+1})$.
The input is fed into the device described in Section~\ref{TWOONE}.
The purpose of this front-end \DLM\ is to transform the information contained in
two-dimensional input vectors (of which only one is present for any given input event),
into a four-dimensional unit vector.
The four-dimensional internal vector of this device
is split into two groups of two-dimensional vectors
$(\widetilde x_{1,n+1},\widetilde x_{4,n+1})$ and $(\widetilde x_{3,n+1},\widetilde x_{2,n+1})$
and each of these two-dimensional vectors is rotated by $45^\circ$.
Put differently, the four-dimensional vector is rotated
once in the (1,4)-plane about $45^\circ$ and once in the (3,2) plane about $45^\circ$.
The order of the rotations is irrelevant.
The resulting four-dimensional vector is then sent
to the input of a second $K=4$ \DLM.
This back-end \DLM\ sends $(x_{1,n+1},x_{2,n+1})/\sqrt{x_{1,n+1}^2+x_{2,n+1}^2}$ through output
channel 0 if it used rule $j=1,2$ (see Eq.~\Eq{HYP1}) to update its internal state.
Otherwise it sends $(x_{3,n+1},x_{4,n+1})/\sqrt{x_{3,n+1}^2+x_{4,n+1}^2}$ through output channel 1.

The operation of the network depicted in Fig.\ref{figbs}
can be analyzed analytically if we disregard transient effects
and assume that the information carried by
events on channel 0 (1) is given by
${\bf y}_{n+1}={\bf y}=(y_{1},y_{2})$
(${\bf y}^\prime_{n+1}={\bf y}^\prime=(y_{3},y_{4})$).
We denote by $p$ the number of events on input channel 0
divided by the total number of events.
Then, the number of events on input channel 1 is given by $1-p$.

In the stationary regime, the internal state
$(\widetilde x_{1,n+1},\widetilde x_{2,n+1},\widetilde x_{3,n+1},\widetilde x_{4,n+1})$
of the front-end \DLM\ (see Fig.\ref{figbs})
learns $(w_{1},w_{2},w_{3},w_{4})=(y_{1}\sqrt{p},y_{2}\sqrt{p},y_{3}\sqrt{1-p},y_{4}\sqrt{1-p})$.
Carrying out the two plane rotations of $45^\circ$
we see that the back-end \DLM\ receives as input the four-dimensional vector
$(w_{1}-w_{4},w_{3}+w_{2},w_{3}-w_{2},w_{1}+w_{4})/\sqrt{2}$.
In the stationary regime, the internal vector
$(x_{1,n+1}, x_{2,n+1}, x_{3,n+1}, x_{4,n+1})$ of the back-end \DLM\
oscillates about
$(w_{1}-w_{4},w_{3}+w_{2},w_{3}-w_{2},w_{1}+w_{4})/\sqrt{2}$.
Therefore, in the stationary regime and for fixed two-dimensional
vectors on input channels 0 and 1, the input-output relation
of the BS network of Fig.~\ref{one-bs} can be written as

\begin{eqnarray}
\left(
\begin{array}{c}
w_1\\
w_2\\
w_3\\
w_4
\end{array}
\right)
\overset{BS}{\longrightarrow}
\frac{1}{\sqrt{2}}
\left(
\begin{array}{c}
w_1-w_4\\
w_3+w_2\\
w_3-w_2\\
w_1+w_4
\end{array}
\right).
\label{BS1}
\end{eqnarray}
Using two complex numbers instead of four real numbers
Eq.~\Eq{BS1} can also be written as

\begin{eqnarray}
\left(
\begin{array}{c}
w_1+i w_2\\
w_3+i w_4
\end{array}
\right)
\overset{BS}{\longrightarrow}
\frac{1}{\sqrt{2}}
\left(
\begin{array}{c}
w_1-w_4+i(w_3+w_2)\\
w_3-w_2+i(w_1+w_4)
\end{array}
\right).
\label{BS2}
\end{eqnarray}

In quantum theory~\cite{QuantumTheory} the presence of photons in the input modes 0 or 1 is represented
by the probability amplitudes ($a_0,a_1)$~\cite{BAYM74,GRAN86,RARI97,NIEL00}.
According to quantum theory~\cite{QuantumTheory}, the probability amplitudes ($b_0,b_1)$
of the photons in the output modes 0 and 1 of a beam splitter are given by~\cite{BAYM74,GRAN86,RARI97,NIEL00}

\begin{eqnarray}
\left(
\begin{array}{c}
b_0\\
b_1
\end{array}
\right)
=
\left(
\begin{array}{c}
a_0+ia_1\\
a_1+ia_0
\end{array}
\right)
=
\frac{1}{\sqrt{2}}
\left(
\begin{array}{cc}
1&i\\
i&1
\end{array}
\right)
\left(
\begin{array}{c}
a_0\\
a_1
\end{array}
\right).
\label{BS3}
\end{eqnarray}
Identifying $a_0$ with $w_1+iw_2=(y_1+iy_2)p$
and $a_1$ with $w_3+iw_4=(y_3+iy_4)(1-p)$ it is clear that by construction,
the \DLM\ network in Fig.~\ref{figbs} will
allow us to simulate a beam splitter,
not by calculating the amplitudes Eq.~\Eq{BS3}
but by a deterministic event-by-event simulation.

In Fig.~\ref{one-bs} (right) we present results of discrete-event simulations using
the \DLM\ network depicted in Fig.~\ref{figbs} (left).
Before the simulation starts,
the internal vectors of the \DLMS\ are given a random value (on the unit sphere).
Each data point represents 10000 events.
All these simulations were carried out with $\alpha=0.99$.
For each set of 10000 events, a uniform
random number in the range $[0,360]$ generates two angles $\psi_0$ and $\psi_1$.
Input channel 0 receives $(y_{1,n+1},y_{2,n+1})=(\cos\psi_0,\sin\psi_0)$
with probability $p_0$.
Input channel 1 receives $(y_{3,n+1},y_{4,n+1})=(\cos\psi_1,\sin\psi_1)$ with probability
$p_1=1-p_0$.
Random processes only enter in the procedure to generate the input data.
The \DLM\ network processes the events sequentially and deterministically.
From Fig.~\ref{one-bs} it is clear that the output of the deterministic
\DLM-based beam splitter reproduces the probability distributions as obtained from
quantum theory~\cite{QuantumTheory}.

\begin{figure*}[t]
\begin{center}
\setlength{\unitlength}{1cm}
\begin{picture}(14,14)
\put(-1,7){\includegraphics[width=10cm]{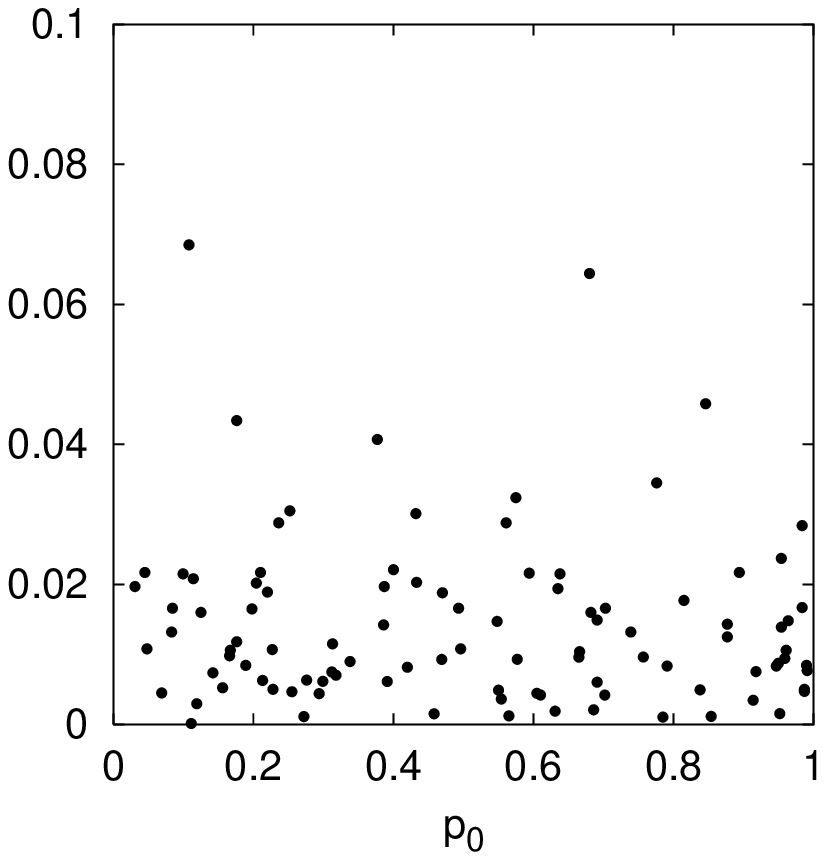}}
\put(7,7){\includegraphics[width=10cm]{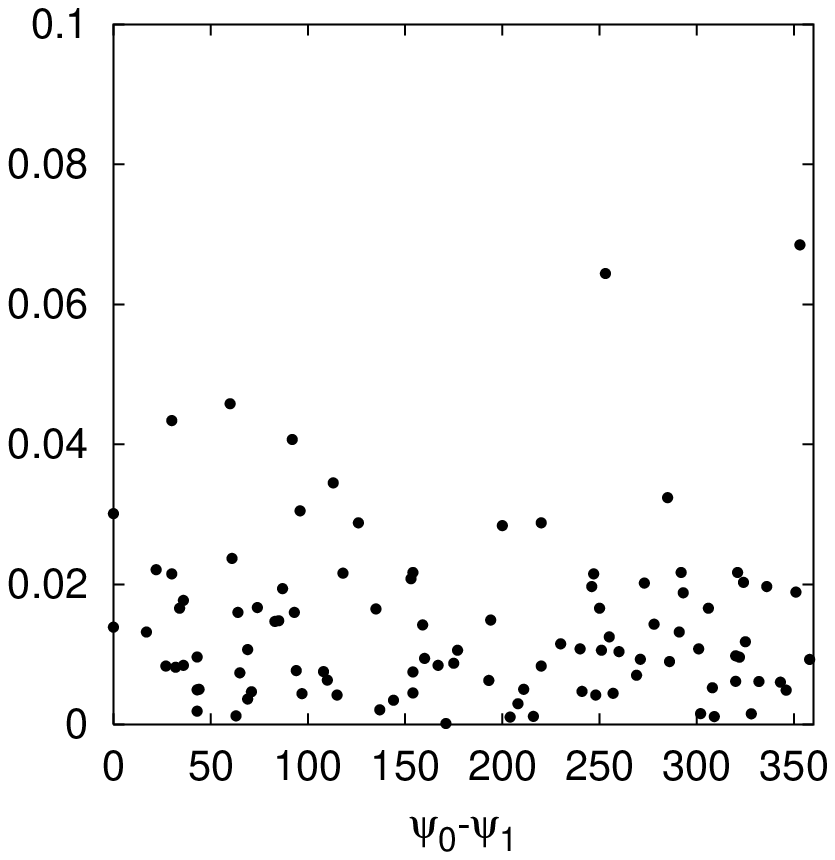}}
\put(-1,0){\includegraphics[width=10cm]{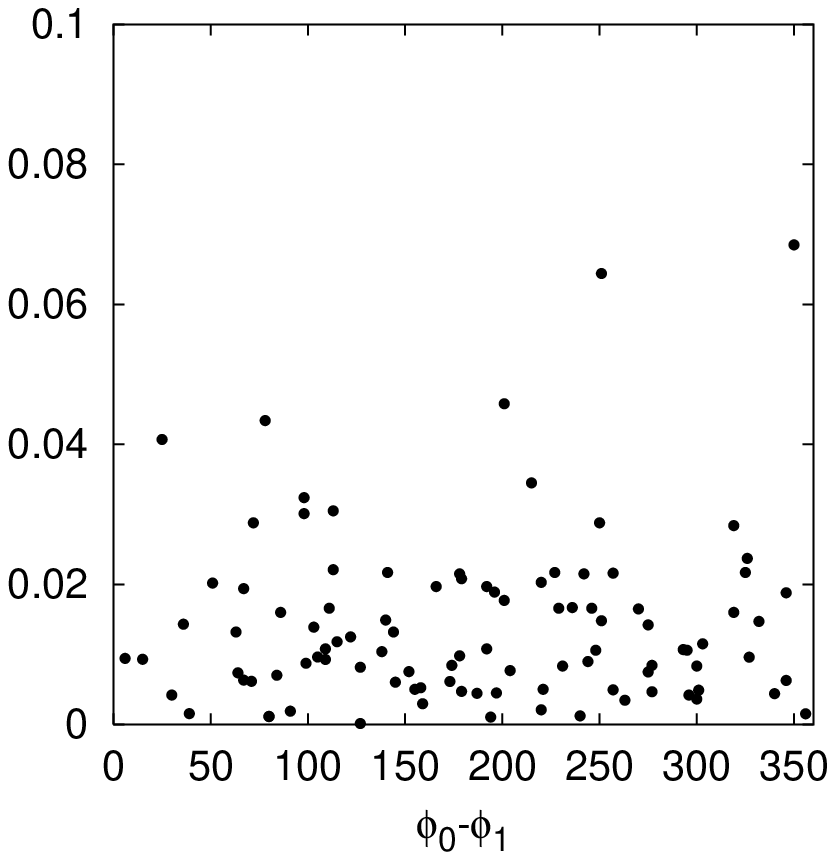}}
\put(7,0){\includegraphics[width=10cm]{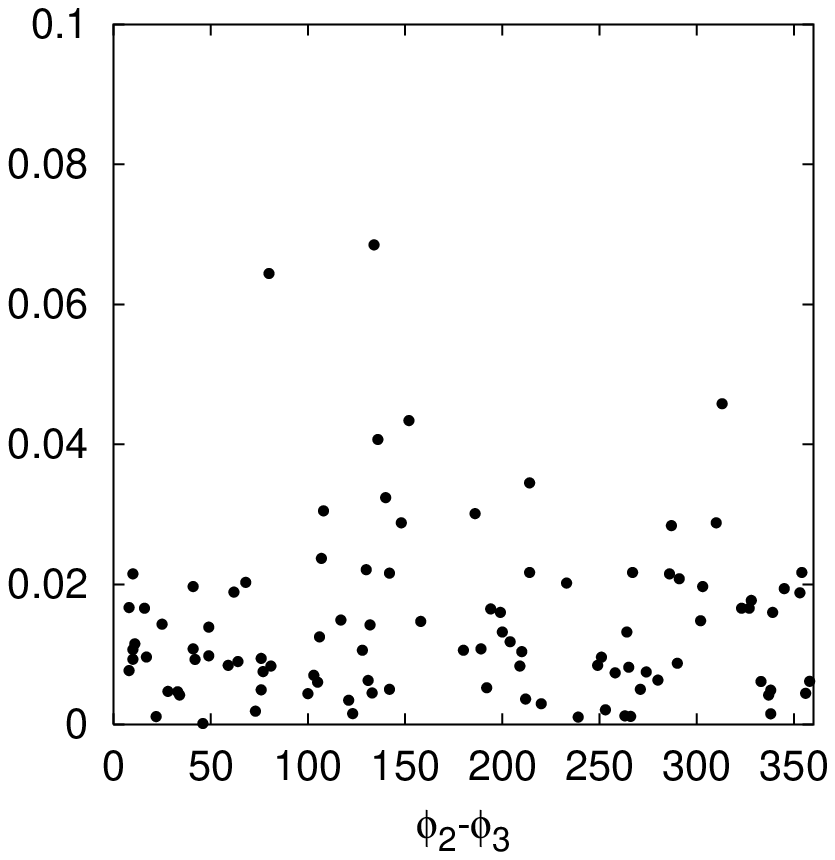}}
\end{picture}
\caption{Absolute value of the difference between
the normalized intensity $N_4/(N_4+N_5)$ in output channel 0
of the event-based \DLM\ simulation
and the result of quantum theory~\cite{QuantumTheory}
for the system of two chained Mach-Zehnder interferometers shown in Fig.~\ref{fig2mz}~\cite{MZIdemo}.
Input channel 0 receives $(y_{1,n+1},y_{2,n+1})=(\cos\psi_0,\sin\psi_0)$
with probability $p_0$. Input channel 1 receives
$(y_{3,n+1},y_{4,n+1})=(\cos\psi_1,\sin\psi_1)$ with probability $1-p_0$.
For each event a uniform random number in the range $[0,360]$ determines $\psi_0$
or $\psi_1$.
Each data point represents a simulation of 10000 events ($N_0+N_1=N_2+N_3=N_4+N_5=10000$).
Top-left: Difference as a function of $p_0$;
Top-right: Difference as a function of $\psi_0-\psi_1$;
Bottom-left: Difference as a function of $\phi_0-\phi_1$;
Bottom-right: Difference as a function of $\phi_2-\phi_3$.
}
\label{two-mz}
\end{center}
\end{figure*}

\begin{figure*}[t]
\begin{center}
\setlength{\unitlength}{1cm}
\begin{picture}(14,14)
\put(-1,7){\includegraphics[width=10cm]{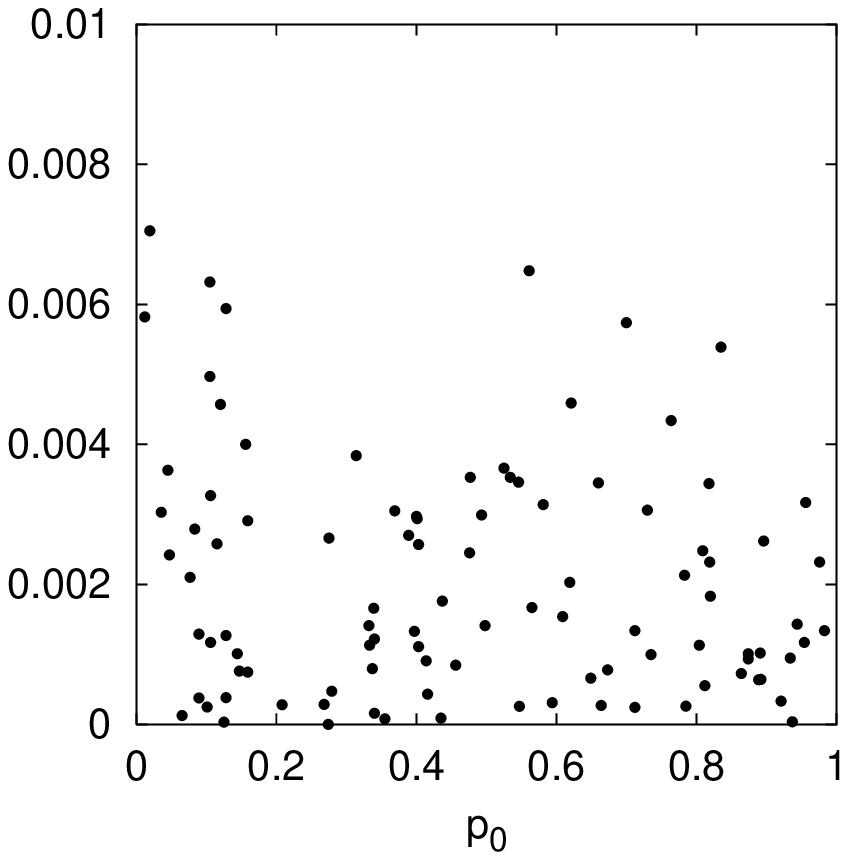}}
\put(7,7){\includegraphics[width=10cm]{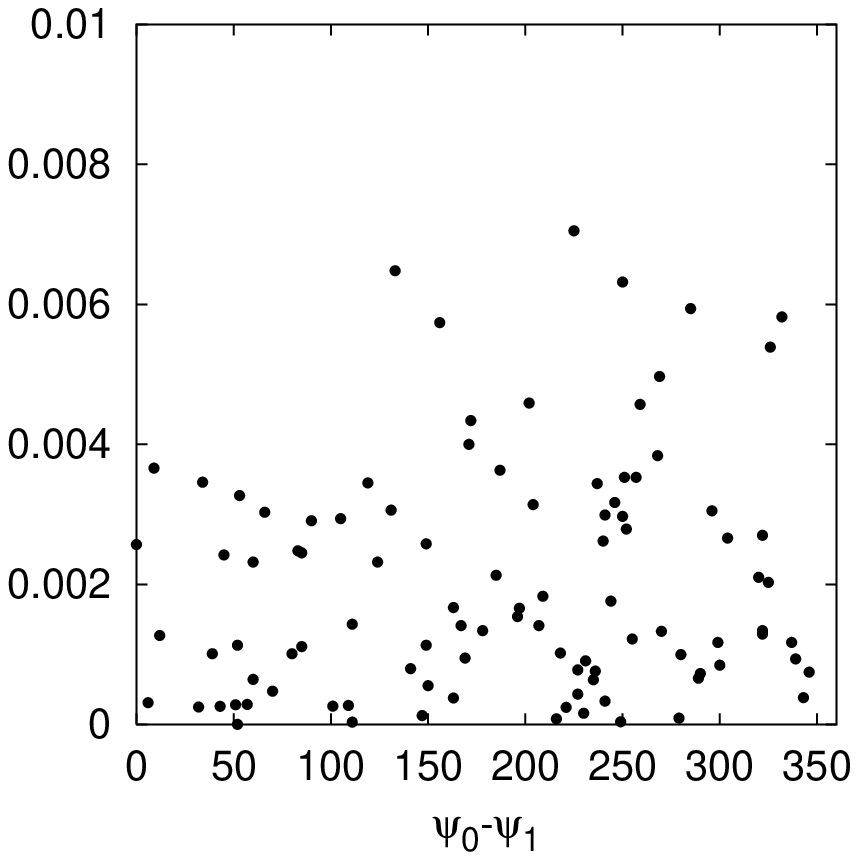}}
\put(-1,0){\includegraphics[width=10cm]{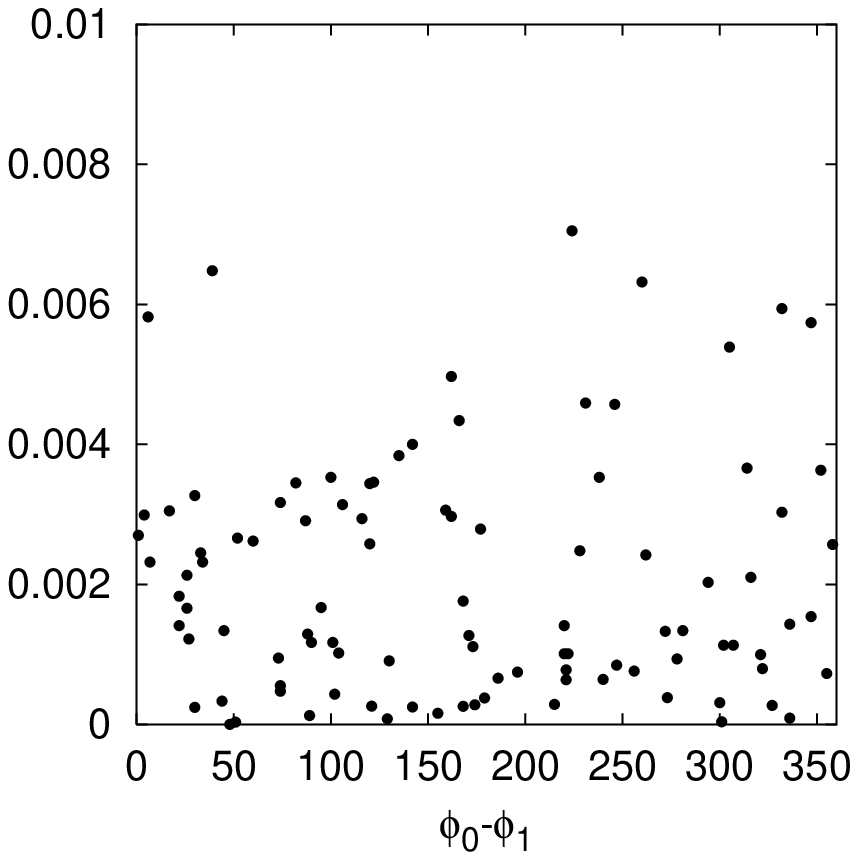}}
\put(7,0){\includegraphics[width=10cm]{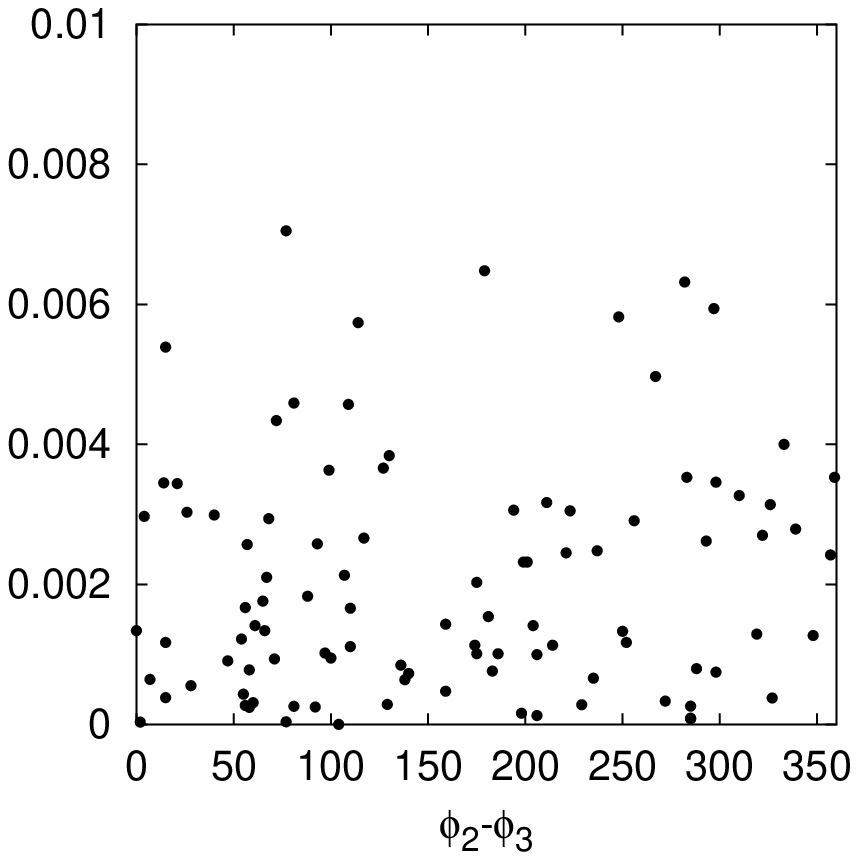}}
\end{picture}
\caption{Same as Fig.~\ref{two-mz} except that
$\alpha=0.9999$ (instead of $\alpha=0.99$)
and
that the 1000000 events (instead of 10000)
per data point were processed by the \DLM\ network
depicted in Fig.~\ref{fig2mz}.
}
\label{two-mz-b}
\end{center}
\end{figure*}

\subsection{Mach-Zehnder interferometer}\label{MZI}

In quantum physics~\cite{QuantumTheory}, single-photon experiments with one beam splitter
provide direct evidence for the particle-like behavior of photons~\cite{GRAN86,HOME97}.
The wave mechanical character appears when one performs single-particle
interference experiments.
In this subsection we construct a \DLM\ network that displays the same
interference patterns as those observed in single-photon
Mach-Zehnder interferometer experiments~\cite{GRAN86}.

The schematic layout of the \DLM\ network is shown in Fig.~\ref{figmz}.
Not surprisingly, it is exactly the same as that of a real Mach-Zehnder interferometer.
The BS network described in the previous subsection
is used for the beam splitters.
The phase shift is taken care of by a passive device that performs a plane rotation.
Clearly there is a one-to-one mapping
from each relevant component in the interferometer
to a processing unit in the \DLM\ network.
Recall that the processing units in the \DLM\ network
only communicate with each other through the message (photon)
that propagates through the network.

According to quantum theory~\cite{QuantumTheory}, the probability amplitudes ($b_0,b_1)$
of the photons in the output modes 0 ($N_2$) and 1 ($N_3$) of the
Mach-Zehnder interferometer are given by~\cite{BAYM74,GRAN86,RARI97,NIEL00}

\begin{eqnarray}
\left(
\begin{array}{c}
b_0\\
b_1
\end{array}
\right)
=
\frac{1}{2}
\left(
\begin{array}{cc}
1&i\\
i&1
\end{array}
\right)
\left(
\begin{array}{cc}
e^{i\phi_0}&0\\
0&e^{i\phi_1}
\end{array}
\right)
\left(
\begin{array}{cc}
1&i\\
i&1
\end{array}
\right)
\left(
\begin{array}{c}
a_0\\
a_1
\end{array}
\right)
.
\label{MZ1}
\end{eqnarray}
Note that in a quantum mechanical setting it is impossible to
simultaneously measure ($N_0/(N_0+N_1)$, $N_1/(N_0+N_1)$)
and ($N_2/(N_0+N_1)$, $N_3/(N_0+N_1)$):
Photon detectors operate by absorbing photons.
However, in our deterministic, event-based simulation there is no such problem.

In Fig.~\ref{one-mz} we present a small selection of simulation results
for the Mach-Zehnder interferometer built from \DLMS.
We assume that input channel 0 receives
$(y_{1,n+1},y_{2,n+1})=(\cos\psi_0,\sin\psi_0)$ with probability
one and that input channel 1 receives no events.
This corresponds to $(a_0,a_1)=(\cos\psi_0+i\sin\psi_0,0)$.
We use uniform random numbers to determine $\psi_0$.
In all these simulations $\alpha=0.99$.
The data points are the simulation results for the
normalized intensity $N_i/(N_0+N_1)$ for i=0,2,3
as a function of $\phi=\phi_0-\phi_1$.
Lines represent the corresponding results of quantum theory~\cite{QuantumTheory}.
From Fig.~\ref{one-mz} it is clear that quantum theory
provides an excellent description of the
deterministic, event-based processing by the \DLM\ network.

The examples presented in Fig.~\ref{one-mz} do not rule
out that there may be settings for the angles
$\psi_0$, $\phi_0$ and $\phi_1$ for which quantum
theory fails to give a good description of the behavior of the \DLM\ network.
However extensive series of simulations show that this is not the case.
Instead of presenting the results of these simulations
we will demonstrate that quantum theory~\cite{QuantumTheory} also
describes the stationary-state input-output behavior of
more extended \DLM\ networks.

As an example we consider the \DLM\ network depicted in Fig.~\ref{fig2mz}.
Obviously this network maps exactly onto two chained Mach-Zehnder interferometers~\cite{MZIdemo}.
Now there are seven parameters $p_0$, $\psi_0$, $\psi_1$, $\phi_0$,
$\phi_1$, $\phi_2$, and $\phi_3$
that may be varied, so simply plotting selected cases is not the proper procedure
to establish that quantum theory describes the stationary-state behavior
of the \DLM\ network.
Therefore we adopt the following strategy.
For each set of 10000 events, we use seven random numbers to fix the
parameters $p_0$, $\psi_0$, $\psi_1$, $\phi_0$, $\phi_1$, $\phi_2$, and $\phi_3$.
Then we collect the data for these 10000 events and compare
the intensity in output channel 0 ($N_4$) and 1 ($N_5$) with the corresponding
results of quantum theory~\cite{QuantumTheory}.
The latter is given by

\begin{eqnarray}
\left(
\begin{array}{c}
b_0\\
b_1
\end{array}
\right)
=
\frac{1}{2\sqrt{2}}
\left(
\begin{array}{cc}
1&i\\
i&1
\end{array}
\right)
\left(
\begin{array}{cc}
e^{i\phi_2}&0\\
0&e^{i\phi_3}
\end{array}
\right)
\left(
\begin{array}{cc}
1&i\\
i&1
\end{array}
\right)
\left(
\begin{array}{cc}
e^{i\phi_0}&0\\
0&e^{i\phi_1}
\end{array}
\right)
\left(
\begin{array}{cc}
1&i\\
i&1
\end{array}
\right)
\left(
\begin{array}{c}
a_0\\
a_1
\end{array}
\right)
.
\label{MZ2}
\end{eqnarray}
For each choice of $\{p_0,\psi_0,\psi_1,\phi_0,\phi_1,\phi_2,\phi_3\}$
we compute the differences
$||b_0|^2-N_4/(N_4+N_5)|$ and $||b_1|^2-N_5/(N_4+N_5)|$.
$N_4$ ($N_5$) is the number of events in the output channel 0 (1)
of the third beam splitter.
$N_0+N_1=N_2+N_3=N_4+N_5$ is the total number of events (10000 in this case).
In Fig.~\ref{two-mz} we show $||b_0|^2-N_4/(N_4+N_5)|$ as a function of
$p_0$,
$\psi_0-\psi_1$,
$\phi_0-\phi_1$,
and
$\phi_2-\phi_3$.
In all these simulations $\alpha=0.99$.
Once again it is clear that quantum theory~\cite{QuantumTheory} provides
a very good description of a \DLM-based simulation
of two chained Mach-Zehnder interferometers.

\subsection{Technical note}

All simulations that we presented in this section have been performed
for $\alpha=0.99$.
From the description of the learning process it is
clear that $\alpha$ controls the rate of learning or, equivalently,
the rate at which learned information can be forgotten.
Furthermore it is evident that the difference between
a constant input to a \DLM\ and the learned value of its
internal variable cannot be smaller than $1-\alpha$.
In other words, $\alpha$ also limits the precision with
which the internal variable can represent a sequence
of constant input values.
On the other hand, the number of events has to balance
the rate at which the \DLM\ can forget a learned input value.
The smaller $1-\alpha$ is, the larger the number of events
has to be for the \DLM\ to adapt to changes in the input data.

We use the last example of Section~\ref{MZI} to illustrate the effect of changing
$\alpha$ and the total number of events $N$.
In Fig.~\ref{two-mz-b} we show the results of repeating the procedure used
to obtain the data shown in Fig.~\ref{two-mz} but
instead of $\alpha=0.99$ and $N=10000$ events per data point,
we used $\alpha=0.9999$ and $N=1000000$ event per data point.
As expected, the difference between the simulation data and the results of
quantum theory decreases if $1-\alpha$ decreases and $N$ increases accordingly.
Comparing Fig.~\ref{two-mz} with Fig.~\ref{two-mz-b} it is clear
that the decrease of this difference is roughly proportional
to the inverse of the square root of the number of events.
Note that each data point in Fig.~\ref{two-mz}
is generated without the use of random processes.

\setlength{\unitlength}{1cm}
\begin{figure*}[t]
\begin{center}
\setlength{\unitlength}{1cm}
\includegraphics[width=12cm]{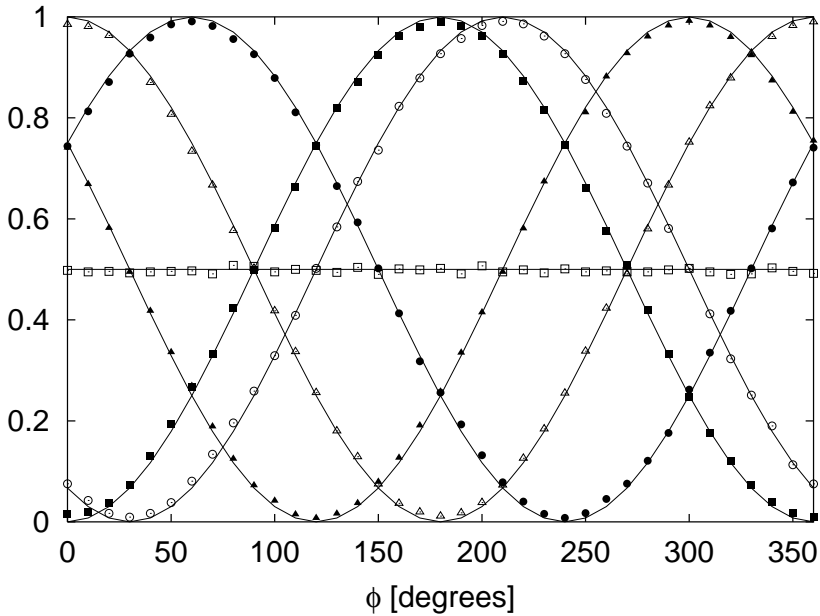}
\caption{
Simulation results for a Mach-Zehnder interferometer built from SLMs
instead of \DLMS.
Each beam splitter sends messages over its output channels 0 and 1
in a random manner.
The simulation procedure and annotations are exactly the same as in Fig.~\ref{one-mz}.
}
\label{one-mz-random}
\end{center}
\end{figure*}

\section{Stochastic learning machines}\label{SLM}

In the stationary regime, the sequence of messages that a \DLM\ (network)
generates is strictly deterministic.
For some applications, e.g. for quantum physics~\cite{QuantumTheory}, it may be desirable
to randomize these sequences.
A marginal modification turns a \DLM\ into a stochastic learning machine (SLM).
Here the term {\sl stochastic} does not refer to the learning process but to
the method that is used to select the output channel that will carry the outgoing
message.

In the stationary regime the components of the internal vector represent
the probability amplitudes.
Comparing the (sums of) squares of these amplitudes with a uniform
random number $0<r<1$ gives the probability for sending the message
over the corresponding output channel.
For instance, in the case of the beam splitter BS (see Fig.~\ref{figbs})
we replace the back-end DLM by a SLM.
This SLM will send a message over output channel 0
if $x^2_{1,n+1}+x^2_{2,n+1}\le r$. Otherwise it will activate output channel 1.
Although the learning process of this modified BS network is still
deterministic, in the stationary regime the output messages are
randomly distributed over the two output channels.
Of course, the distribution of output messages is the same as that of the original \DLM-network.

Replacing \DLMS\ by SLMs in a \DLM-network changes the order in which
messages are being processes by the network but leaves the content of the
messages intact.
Therefore, in the stationary regime, the distribution of messages
over the outputs of the SLM-network is essentially the same as that
of the original DLM network.

As an illustration of the use of SLMs, we replace the two back-end \DLMS\ in the
Mach-Zehnder interferometer network (see Fig.~\ref{figmz} (left))
by their ``randomized'' version and repeat the procedure
that generates the data of Fig.~\ref{figmz} (right).
The results of these simulations are shown in Fig.~\ref{one-mz-random}.
Not unexpectedly, the randomness in the output channel selection
is reflected by a (small) increase of the scatter on the data points.
In this simulation, the output channels 0 and 1 of each beam splitter are activated
in a random manner and the functional dependence of
$N_0/(N_0+N_1)$, $N_1/(N_0+N_1)$,
$N_2/(N_2+N_3)=N_2/(N_0+N_1)$ and $N_3/(N_2+N_3)$ on $\phi$
is still in full agreement with quantum theory~\cite{QuantumTheory}.
In other words, this SLM-network performs a genuine, event-by-event
simulation of the ideal (perfect detectors, etc.) version
of both the single-photon beam splitter and Mach-Zehnder interferometer experiments
by Grangier et al~\cite{GRAN86}.

\section{Discussion}\label{SUMM}

We have proposed a new procedure to construct deterministic algorithms
that have primitive learning capabilities. 
We have used these algorithms to build deterministic learning machines (DLMs). 
A DLM learns by processing event after event but does not store the data contained
in an individual event.
Connecting the input of a DLM to the output of another DLM yields
a locally connected network of DLMs. 
A DLM within the network locally processes the information contained
in an event and responds by sending a message that may be used as input for another DLM.
A distinct feature of a DLM network is that at any given time,
only one event (message) is propagating through the network. 
The DLMs process messages in a sequential manner and
only communicate with each other by message passing. 

We have demonstrated that DLM networks can discover relationships
between successive events (see Section~\ref{NDIM}) and that
certain classes of DLM networks exhibit behavior that is usually
only attributed to quantum systems.
In Sections~\ref{QI} and \ref{SLM} we have presented DLM networks
that simulate quantum interference on an event-by-event basis.
More specifically, we map each physical part of the real Mach-Zehnder interferometer
onto a DLM and the messages (phase shifts in this case) are carried by photons.
No ingredient other than simple geometry is used to specify the update rules of the DLMs.
 
As the network processes event after event, the network generates output events
that build an interference pattern that is described by the quantum theory~\cite{QuantumTheory}
of the single-photon beam splitter and Mach-Zehnder interferometer.
To illustrate that DLM networks are indeed capable of simulating quantum interference
on an event-by-event basis we also simulate an experiment involving
three beam splitters (i.e. two chained Mach-Zehnder interferometers) and demonstrate
that quantum theory~\cite{QuantumTheory} also describes the behavior of this network.

The results presented in Sections~\ref{QI} and \ref{SLM} suggest that we may have discovered
a systematic procedure to construct algorithms that simulate quantum phenomena
using deterministic, local, and event-by-event-based processes.
We emphasize that our approach is not a proposal for another
interpretation of quantum mechanics.
Our approach is not an extension of quantum theory in any sense:
The probability distributions of quantum theory appear as the result of a deterministic,
causal learning process, and not vice versa (see Section~\ref{QI})~\cite{PENR90}.
Our results suggest that quantum mechanical behavior may originate from
an underlying deterministic process~\cite{tHOO01,tHOO02}.
Indeed, it is somewhat ironic that in order to mimic the apparent randomness with which
events are observed in experiments, we have to explicitly randomize the output of the DLMs
to mask the underlying deterministic processes (see Section~\ref{SLM}).
To the best of our knowledge, this paper contains the first demonstration
that quantum interference can be simulated on an event-by-event basis
using local, causal, and deterministic processes, and
without using concepts such as wave fields or particle-wave duality.

At this point it may be worthwhile to recall what a DLM actually does.
In a simple physical picture, a DLM is a device (e.g. beam splitter, polarizer) that
exchanges information with the particles that pass through it.
The DLM tries to do this in an effective manner.
It learns by comparing the message carried by an event with predictions
based on the knowledge acquired by the DLM during the processing of previous events.
Effectively this comparison amounts to a minimization of the squared error
(see Section~\ref{ILLU}).
Schr{\"o}dinger used exactly the same principle to derive his famous equation~\cite{SCHR26a}
but called this approach ``unverst\"andlich'' in a subsequent publication~\cite{SCHR26b}.

In a future publication we will show that the approach introduced in this paper
can be employed to perform event-based simulations
of a universal quantum computer~\cite{KRIS04}.
It has been shown that the time evolution of the wave function
of a quantum system can be simulated on a quantum computer~\cite{ZALK98,NIEL00}.
Therefore it should be possible to compute the real-time dynamics
of these systems (including the double-slit experiment mentioned in the introduction)
through discrete-event simulation by constructing appropriate DLM networks.

\section*{Acknowledgement}
We thank S. Miyashita for extensive discussions.



\raggedright



\begin{thebibliography}{999}
\bibitem{PerfectExperiments}
In this paper we disregard
limitations of real experiments such as detector efficiency,
imperfection of the source, biprism etc.
\bibitem{QuantumTheory}
We make a distinction between quantum theory and quantum physics.
We use the term {\sl quantum theory} when we refer to
the mathematical formalism, i.e., the postulates
of quantum mechanics (with or without the wave function
collapse postulate)~\cite{BALL03} and the rules (algorithms) to compute the
wave function.
The term {\sl quantum physics} is used for microscopic, experimentally observable phenomena
that do not find an explanation within the mathematical
framework of classical mechanics.
\end{thebibliography}
\end{document}